\documentclass[referee]{aa}
\pdfoutput=1
\usepackage{float,lscape}
\usepackage{txfonts}
\usepackage{natbib}
\usepackage{hyperref, mathtools}
\usepackage{graphicx}
\DeclareGraphicsExtensions{.pdf,.png,.jpg}
\bibpunct{(}{)}{;}{a}{}{,}

\newcommand{\mass}{\mbox{$m_{\star}$}}

\newcommand{\Mabs}{\mbox{$M_{\rm abs}$}}

\newcommand{\feh}{\mbox{$\rm [{\rm Fe}/{\rm H}]$}}
\newcommand{\mh}{\mbox{$\rm [{\rm M}/{\rm H}]$}}
\newcommand{\afe}{\mbox{$\rm [{\alpha}/{\rm Fe}]$}}

\newcommand{\Teff}{\mbox{$T_{\rm eff}$}}

\newcommand{\beq}{\begin{equation}}
\newcommand{\eeq}{\end{equation}}
\newcommand{\beqa}{\begin{eqnarray}}
\newcommand{\eeqa}{\end{eqnarray}}

\begin{document}
\title{Spectro-photometric distances to stars: \\
A general-purpose Bayesian approach}
\author{
Bas\'{i}lio X. Santiago\inst{1,2} \and 
Doroth\'{e}e E. Brauer\inst{3} \and 
Friedrich Anders\inst{3,2} \and 
Cristina Chiappini\inst{3,2} \and 
Anna B. Queiroz\inst{1,2} \and \\
L\'{e}o Girardi\inst{4,2} \and 
Helio J. Rocha-Pinto\inst{5,2} \and 
Eduardo Balbinot\inst{1,2} \and 
Luiz N. da Costa\inst{6,2} \and 
Marcio A.G. Maia\inst{6,2} \and \\
Mathias Schultheis\inst{7} \and 
Matthias Steinmetz\inst{3} \and 
Andrea Miglio\inst{8} \and 
Josefina Montalb\'{a}n\inst{9} \and 
Donald P. Schneider\inst{10,11} \and \\
Timothy C. Beers\inst{12} \and 
Peter M. Frinchaboy\inst{13} \and 
Young Sun Lee\inst{14} \and 
Gail Zasowski\inst{15}
}
\authorrunning{Santiago, Brauer, Anders, Chiappini, Queiroz et al.}    
\titlerunning{Spectro-photometric distances to stars}
\institute{
Instituto de F\'{i}sica, Universidade Federal do Rio Grande do Sul, Caixa Postal 15051, Porto Alegre, RS - 91501-970, Brazil 
\and
Laborat\'{o}rio Interinstitucional de e-Astronomia - LIneA, Rua Gal. Jos\'{e} Cristino 77, Rio de Janeiro, RJ - 20921-400, Brazil 
\and
Leibniz-Institut f\"{u}r Astrophysik Potsdam (AIP), An der Sternwarte 16, 14482, Potsdam, Germany 
\and
Osservatorio Astronomico di Padova - INAF, Vicolo dell'Osservatorio 5, I - 35122 Padova, Italy 
\and
Universidade Federal do Rio de Janeiro, Observat\'{o}rio do Valongo, Ladeira do Pedro Ant\^{o}nio 43, 20080-090 Rio de Janeiro, Brazil \and
Observat\'{o}rio Nacional, Rua Gal. Jos\'{e} Cristino 77, Rio de Janeiro, RJ - 20921-400, Brazil 
\and
Observatoire de la Cote d'Azur, Laboratoire Lagrange, CNRS UMR 7923, B.P. 4229, 06304 Nice Cedex, France 
\and
School of Physics and Astronomy, University of Birmingham, Edgbaston, Birmingham, B15 2TT, United Kingdom 
\and
Institut d'Astrophysique et de G\'{e}ophysique, All\'{e}e du 6 ao\^{u}t, 17 - B\^{a}t. B5c, B-4000 Li\`{e}ge 1 (Sart-Tilman), Belgium 
\and
Department of Astronomy and Astrophysics, The Pennsylvania State University, University Park, PA 16802 
\and
Institute for Gravitation and the Cosmos, The Pennsylvania State University, University Park, PA 16802 
\and
Department of Physics and JINA Center for the Evolution of the Elements, University of Notre Dame, Notre Dame, IN 46556, USA
\and
Department of Physics \& Astronomy, Texas Christian University, TCU Box 298840, Fort Worth, TX 76129 
\and
Department of Astronomy and Space Science, Chungnam National University, Daejeon 34134, Republic of Korea
\and
Dept of Physics and Astronomy, Johns Hopkins University, Baltimore, MD, 21210, USA
}

   \abstract
{Determining distances to individual field stars is a necessary step towards mapping Galactic structure and determining spatial variations in the chemo-dynamical properties of stellar populations in the Milky Way.}
{In order to provide stellar distance estimates for various spectroscopic surveys, we have developed a code that estimates distances to stars using measured spectroscopic and photometric quantities. We employ a Bayesian approach to build the probability distribution function over stellar evolutionary models given these data, delivering estimates of model parameters (including distances) for each star individually. Our method provides several alternative distance estimates for each star in the output, along with their associated uncertainties. This enables the use of our method even in the absence of some measurements.}
{The code was first tested on simulations, successfully recovering input distances to mock stars with $\lesssim 1\%$ bias. We found the uncertainties scale with the uncertainties in the adopted spectro-photometric parameters. The method-intrinsic random distance uncertainties for typical spectroscopic survey measurements amount to around 10\% for dwarf stars and 20\% for giants, and are most sensitive to the quality of $\log g$ measurements.}
{The code was then validated by comparing our distance estimates to parallax measurements from the {\it Hipparcos} mission for nearby stars ($< 300$ pc), to asteroseismic distances of CoRoT red giant stars, and to known distances of well-studied open and globular clusters. The photometric data of these reference samples cover both optical and infrared wavelengths. The spectroscopic parameters are also based on spectra taken at various wavelengths, with varying spectral coverage and resolution: the Sloan Digital Sky Survey programs SEGUE and APOGEE, as well as various ESO instruments.}
{The external comparisons confirm that our distances are subject to very small systematic biases with respect to the fundamental {\it Hipparcos} scale ($+0.4 \%$ for dwarfs, and $+1.6\%$ for giants). The typical random distance scatter is $18\%$ for dwarfs, and $26\%$ for giants. For the CoRoT-APOGEE sample, which spans Galactocentric distances of $4-14$ kpc, the typical random distance scatter is $\simeq 15\%$, both for the nearby and farther data. Our distances are systematically larger than the CoRoT ones by about $+9\%$, which can mostly be attributed to the different choice of priors. 
The comparison to known distances of star clusters from SEGUE and APOGEE has led to significant systematic differences for many cluster stars, but with opposite signs, and with substantial scatter. Finally, we tested our distances against those previously determined for a high-quality sample of giant stars from the RAVE survey, again finding a small systematic trend of $+5\%$ and an rms scatter of $30\%$. 
Efforts are underway to provide our code to the community by running it on a public server.}
{}


\maketitle

\section{Introduction}\label{introd}

A crucial step towards studying stellar populations and their variation across the Galaxy is to measure reliable stellar distances.
The parallax method is currently viable only for very local stars, although the recently launched Gaia satellite \citep{Perryman2001} is expected to dramatically increase the number of parallax and proper motion measurements within a significant portion of the Milky Way. 
Precise distances may also be obtained from asteroseismology, but only for a fraction of stars subject to systematic variability studies \citep{Miglio2012, Rodrigues2014}.
Indirect methods of distance determination, based on photometric
and spectroscopic quantities and their relation to stellar absolute magnitudes, can be used for much more distant stars. Photometric distances have been
applied to multi-band optical and near infra-red data
from different surveys to model the spatial distribution of Galactic stars and to study its substructures \citep[e.g.,][]{Juric2008,Correnti2010,Minniti2011}.
Recent Sloan Digital Sky Survey III \citep[SDSS-III,][]{Eisenstein2011} spectroscopic
surveys, such as the Sloan Extension for Galactic Understanding and Exploration
\citep[SEGUE,][]{Yanny2009} and the Apache Point Observatory 
Galactic Evolution Experiment \citep[APOGEE,][]{AllendePrieto2008} have  
produced many more observational constraints, including stellar 
atmospheric parameters, kinematical and chemical data, that can be used to estimate reliable distances to tens of thousands of stars sampled from the SDSS and 
2MASS \citep{Skrutskie2006} photometric data. These surveys have already had a clear impact on our understanding of the Galaxy 
\citep{Carollo2007,Carollo2010,Lee2011,Schlesinger2012,Carollo2012,Cheng2012,Bovy2012c,Bovy2012a,Anders2014,Hayden2015}. 
The recently completed RAdial Velocity Experiment (RAVE) survey (\citealt{Steinmetz2006}) has also provided atmospheric parameters, radial velocities and chemical abundances for six individual elements for more than 400 000 stars (\citealt{Kordopatis2013}). 
Ongoing spectroscopic surveys, such as the Gaia-ESO Survey (GES; \citealt{Gilmore2012}), the Galactic Archaeology with HERMES survey (GALAH; \citealt{Zucker2012}), or the LAMOST Experiment for Galactic Understanding and Exploration \citep[LEGUE;][]{Deng2012}, are continuously increasing the number of stars with available spectroscopic information.
In fact, these large spectro-photometric data sets will, when analysed together, 
allow a description of the structure and substructures of the 
Galaxy with unprecedented detail, excellent statistics, and full use of the 6D phase-space information.
Phase-space reconstruction of stellar distributions, 
coupled to spectroscopic abundances for large samples, 
will continue to provide challenging quantitative tests to models of Galaxy 
formation and evolution. 

In order to take full advantage of the large set of available spectroscopic and  photometric parameters from recent surveys, probabilistic inference has been
used by several authors to infer ages, absolute magnitudes, extinction, distances, among other parameters \citep[e.g.,][]{Pont2004,Jorgensen2005,Bailer-Jones2011,Serenelli2013,Schoenrich2014}. In the context of stellar distances, a series of
papers with increasing levels of sophistication has been presented by the RAVE
collaboration \citep{Breddels2010,Zwitter2010,Burnett2010}.
\cite{Burnett2010} make use of a comprehensive set of measured parameters
and their estimated uncertainties to infer, for each star, the probability 
distribution that a set of chosen stellar models generate the data. The method was further refined by \cite{Burnett2011} and \cite{Binney2014a}, and later used to study Galactic chemo-dynamics in, e.g.,\citet{Boeche2013, Boeche2013a, Binney2014} or \citet{Kordopatis2015}.

In this paper, we follow a similar theoretical background as those authors,
and implement a code that computes spectro-photometric distances with the goal
of mapping large stellar samples in three dimensions or in phase-space.
We are motivated by the analyses of SDSS-III SEGUE and APOGEE data 
being led by the Brazilian Participation Group (BPG) and the Leibniz-Institut f\"{u}r Astrophysik Potsdam (AIP), which are presented in \cite{Anders2014}
and Brauer (2015; PhD thesis, subm.). These papers use APOGEE giants and SEGUE G-dwarfs, respectively, to improve chemo-dynamical constraints to the Galactic components, especially the discs (see discussions in \citealt{Minchev2014,Chiappini2015b}). 
Since SEGUE and APOGEE targets were selected based on different photometric data
and have different spectral coverage and resolution, our basic challenge is to
ensure that accurate distances are computed for datasets
of vastly different provenance, in a homogeneous way. Our emphasis is therefore on confronting our distance estimates with as many reference samples as possible.
A direct comparison with the RAVE distances obtained by \cite{Binney2014a} for the high-quality giant sample studied in \cite{Boeche2013} is also provided.
In Table \ref{surveytable} we list the main characteristic of the surveys we have applied our method to.

\begin{centering}
\begin{table*}
\caption{Spectroscopic stellar surveys for which distances were computed using the method presented in this paper. For each case we provide basic information, such as duration, number of stars for different signal-to-noise ratios (SNR), spatial and spectral coverage, and spectral resolving power.}
\begin{footnotesize}
\begin{tabular}{lllllll}
\hline \hline
  \multicolumn{1}{c}{Survey} &
  \multicolumn{1}{c}{Year} &
  \multicolumn{1}{c}{$N_{\mathrm{stars}}$} &
  \multicolumn{1}{c}{Spatial coverage} &
  \multicolumn{1}{c}{Typical SNR} &
  \multicolumn{1}{c}{$\lambda$ range} &
  \multicolumn{1}{c}{$R$} \\
\hline
SEGUE (DR9 G dwarfs) & 2004-2008 & $\simeq$ 120,000 (35,000) & $\simeq$ 1,300 deg$^2$ & 25 (35) & $0.38-0.92 \mu$m & 2,000 \\
APOGEE (DR10 HQ giants) & 2011-2014 & $\simeq$ 100,000 (22,000) & $\simeq$ 2,800 deg$^2$& 100 (120) & $1.5-1.7 \mu$m & 22,500 \\
RAVE (DR4 HQ giants) & 2006-2013 & $\simeq$ 500,000 (9,000) & $\simeq$ 20,000 deg$^2$ & 30 (70) & $0.84-0.88 \mu$m & 7,500 \\
\hline
\end{tabular}
\end{footnotesize}
\label{surveytable}
\end{table*}
\end{centering}

The outline of the paper is as follows. In Sec. \ref{method} we review the 
method, introduce our notation, and show the results of initial validation 
tests. An analysis of the performance of our code in terms of internal accuracy and precision is provided. We further discuss how biased stellar parameters and prior assumptions influence the estimated distances. In Sec. \ref{validation} we compare our distances to several previous distance determinations that can be used as a reference, given their higher precision and more controlled systematics. 
We also compare our results for a high-quality sample of RAVE giant stars to those obtained by the RAVE collaboration.
Our summary, conclusions, and future plans are provided in Sec. \ref{conclusion}.

\section {The Method}
\label{method}
\smallskip

The general method adopted for this study 
uses a set of measured photometric and spectroscopic parameters, 
such as metallicity, \feh, alpha element enhancement, \afe, effective temperature, \Teff, surface gravity, $\log g$, intrinsic apparent magnitude, $m$, and colors, 
to estimate the distance to individual stars. 
These quantities are compared to predictions from stellar
evolutionary models. The comparison between model and measured parameters
follows a statistical approach that is similar to previous works \citep{Burnett2010, Burnett2011, Binney2014a}.

In brief, assuming that the errors in the measured parameters follow a normal 
distribution, the probability that a measured value of some quantity, 
$x \pm\sigma_x$, is consistent with some theoretical value, $x_0$, is given by
$$
P(x,\sigma_x | x_0) = \mathcal{N}_{x_0, \sigma}(x) = \frac{1}{\sqrt{2\pi}\ \sigma_{x}}
\exp \left[-\frac{(x-x_0)^2}{2\sigma_x^2}\right]. \eqno{(1)}
$$
We can easily extend this reasoning to a set of (independent) measured parameters, $\vec{x}=\{x_1, ..., x_n\}$, with associated Gaussian uncertainties, ${\vec{\sigma_x}}$, whose theoretical values according to a given 
model are ${\vec{x_0}}$, by writing:
$$
P({\vec{x}},{\vec{\sigma_x}} \vert {\vec{x_0}}) = \prod_{i} \mathcal{N}_{x_{0,i}, \sigma_i}(x_i)
\eqno{(2)}
$$
\noindent where the product is taken over all the measured parameters of a 
single star being confronted
to the model values. The expression above gives the likelihood
of measuring the set $\{\vec{x}, \vec{\sigma_x}\}$ given a model $\vec{x_0}$.
According to Bayes's theorem, we may compute the posterior probability distribution (the probability of the model, given the data) as:

$$ P({\vec{x_0}} \vert {\vec{x}},{\vec{\sigma_x}}) = \frac{{P({\vec{x}},{\vec{\sigma_x}} \vert {\vec{x_0}})~P({\vec{x_0}}) }}{P(\vec{x},\vec{\sigma_x})} , \eqno (3)$$

\noindent where the numerator contains the likelihood and the model prior probability, and the denominator depends only on the measured parameters and their uncertanties.\footnote{Because we are only interested in inferring the best model parameter (in our case the distance) for a specific set of models, this term is dropped in further computations, as it merely represents a constant which can be normalised out \citep[e.g.][Chap. 5]{Ivezic2013}.} 

In order to evaluate the probability of some specific model quantity, $\theta \coloneqq x_{0, i}$, we consider the marginal posterior probability distribution for this quantity, which is obtained by integrating over all variables of Equation (3), except $\theta$:

$$p(\theta) \coloneqq P(\theta|{\vec{x}},{\vec{\sigma_x}}) = \int \mathrm{d}x_{0, 0} ... \mathrm{d}x_{0, i-1} \mathrm{d}_{0, i+1} ... \mathrm{d}x_n
P({\vec{x_0}}\vert {\vec{x}},{\vec{\sigma_x}}) \eqno (4) $$

As mentioned earlier, a typical set of measured parameters includes $\vec{x} = 
\{\mh, \Teff, \log g, \mathrm{colors}, m \}$. As for the stellar models,
besides the theoretical values of the same parameters, they also  
involve other quantities such as mass \mass, age $\tau$, 
and absolute magnitude $\Mabs$. And we take $\theta = d$, where $d$ is
the star's model distance computed as $d~[\mathrm{pc}]=10^{0.2(m-\Mabs+5)}$.

\subsection{Distance Uncertainties} \label{uncerts}

Our code can deliver various statistics for the desirable quantity. In this work, and following \cite{Rodrigues2014}, we compute a star's distance $d$ as the median of the marginalised posterior probability distribution, $p(d)$ (Eq. 4). To estimate the uncertainties, we report the 68\% and 95\% upper and lower credible intervals of the median, $d_{68L}, d_{68U}, d_{95L}$, and $d_{95U}$. When quoting a single uncertainty value, we use the definition 
$$ \sigma(d) \coloneqq 0.5\cdot(d_{68U}-d_{68L}).$$
Whenever there is some ambiguity, we use the subscript ``BPG'' to denote our Bayesian distance estimates\footnote{BPG is short for the SDSS-III Brazilian Participation Group}.

\subsection{A test suite of simulated stars}\label{testset}

\begin{figure}
\begin{center}
\includegraphics[width=\linewidth]{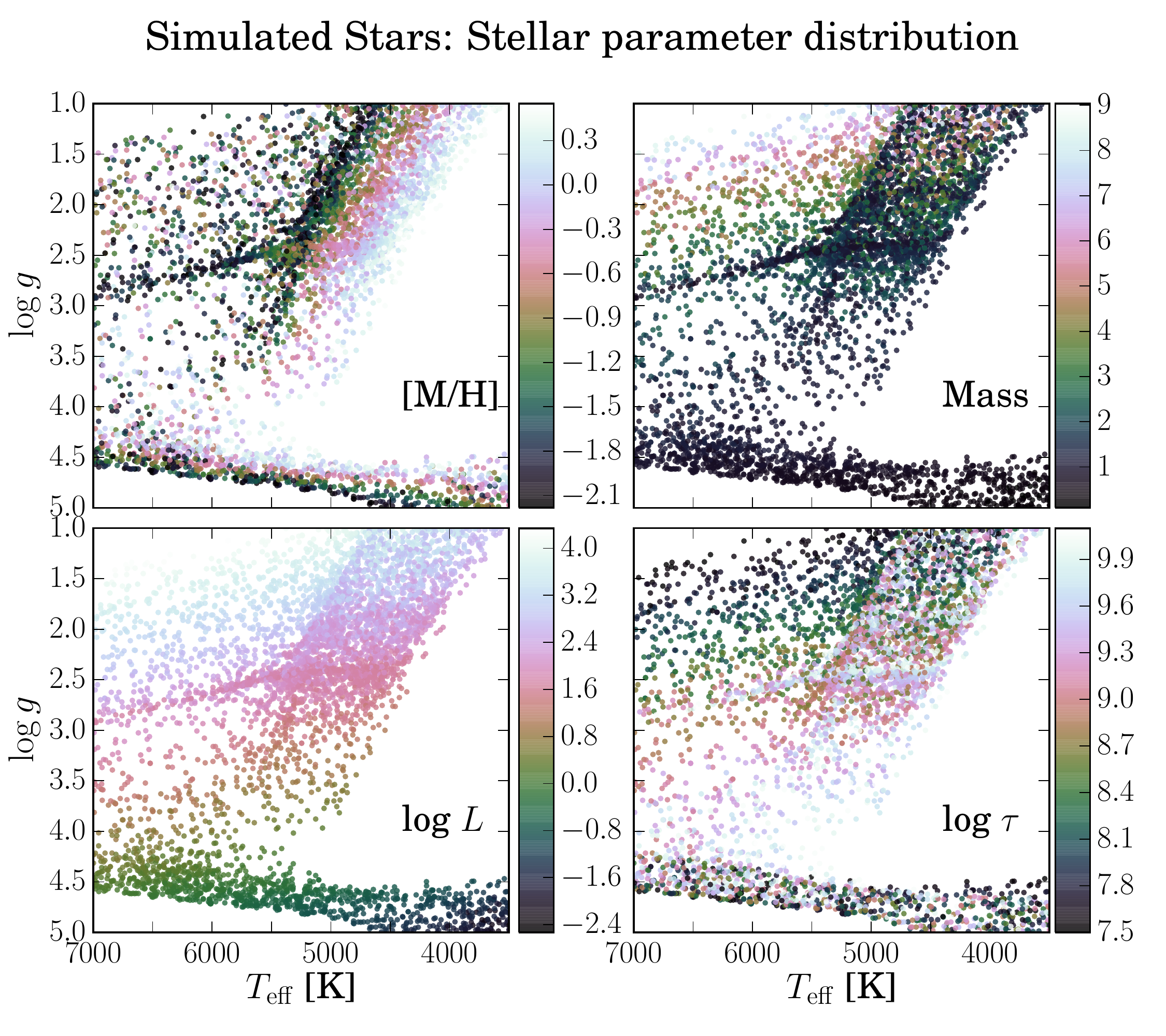}
\caption{\small Stellar parameter distributions of the simulated PARSEC sample in the \Teff--$\log$ g diagram (without observational errors added). {\it Upper left:} solar-scaled metallicity, {\it upper right:} mass (in $M_{\odot}$), {\it lower left:} luminosity (in $L_{\odot}$), {\it lower right:} age (in years).}  
\label{simsample}
\end{center}
\end{figure}

In order to verify the method and our implementation, we use a set of simulated stars drawn from a grid of PARSEC 1.2S stellar models \citep{Bressan2012, Tang2014, Chen2015} with a narrow metallicity step of $\Delta\mh ~ = 0.1$ and spanning the range $-2.2 \leq \mh \leq 0.6$, and age steps of $\log \tau\, [\mathrm{yr}] = 0.05$ in the range $7.5 \leq \log \tau\, [\mathrm{yr}] \geq 10.15$. 
Specifically, we randomly select 5000 models which fall into the parameter regime targeted by large-scale spectroscopic surveys, by restricting model space to $3000<T_{\mathrm{eff}}<7000$ K and $1<\log g<5$ (see Fig. \ref{simsample}). These simulated stars are assigned random distances between 0.1 and 6 kpc, drawn from a uniform distribution. The simulated photometry is in the 2MASS $JHK_s$ system \citep{Cutri2003}, with no extinction applied.

By adding typical (conservative) Gaussian random errors to the model observables, we then create two sets of spectrophotometric ``observations'' from this file: a ``high-resolution-like'' set of spectroscopically measured quantities ($e_{T_{\mathrm{eff}}}=100$ K, $e_{\log g}=0.1$ dex, $e_{\mathrm{[Z/H]}}=0.1$ dex)\footnote{These values can be considered ``typical'' for stellar parameters obtained from high signal-to-noise-ratio (SNR) spectra from a multi-object high-resolution spectrograph like FLAMES/GIRAFFE or APOGEE. For very-high-resolution spectra from, e.g., HARPS or UVES, these values may even be lower by about a factor of 2.}, and a ``low-resolution-like'' set ($e_{T_{\mathrm{eff}}}=200$ K, $e_{\log g}=0.2$ dex, $e_{\mathrm{[Z/H]}}=0.15$ dex). We assume that the errors are uncorrelated and the observational uncertainties are correctly estimated. Photometric errors are also assumed to be Gaussian and of the order of 0.01 mag.

Some of the effects of possible deviations from these assumptions are discussed in the next Sections (stellar parameter biases, incomplete parameter sets, extinction, Galactic density priors). Other effects (e.g., [$\alpha$/Fe]-enhanced isochrones, correlated and non-Gaussian stellar parameter errors) are beyond the scope of this paper, and deferred to a future investigation.

\begin{figure*}
\begin{center}
\includegraphics[width=\textwidth]{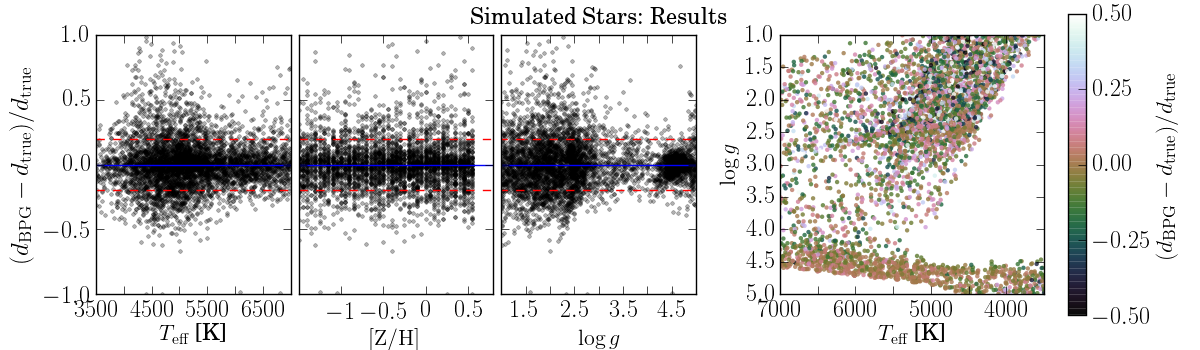}\\
\includegraphics[width=\textwidth]{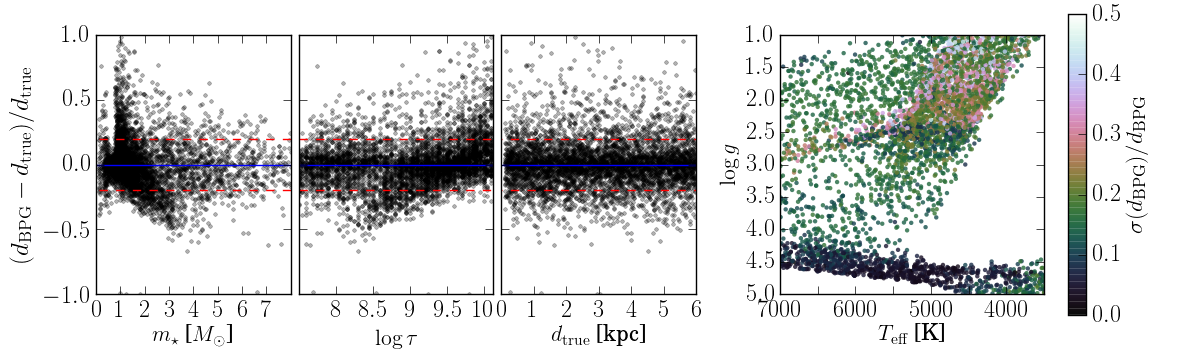}
 \caption{\small Results from the test with the simulated PARSEC sample with ``high-res.-like'' observational errors ($e_{T_{\mathrm{eff}}}$ = 100K, $e_{\mathrm{[Z/H]}}$ = 0.1 dex, $e_{\log g}$ = 0.1 dex, and $e_{m_{\mathrm{obs}}}$ = 0.01 mag.), using the full input parameter set (Set 1: $\vec{x} = \{\mh, \Teff, \log g, m, \mathrm{colors} \}$). 
{\it Left six panels:} Relative distance errors (observed -- true) of the simulated stars shown as a function of the (true) main stellar model parameters \Teff, \mh, $\log g$, mass $m_{\star}$, age $\tau$, and distance $d_{\mathrm{true}}$. The solid line marks the identity line, the dashed lines indicate distance deviations of 20\%. 
{\it Upper right panel:} Relative distance error distribution in the \Teff $-\log g$ diagram. 
{\it Lower right panel:} Relative distance uncertainty distribution in the \Teff $-\log g$ diagram. 
Clearly, distances to giant stars are much more prone to systematics and have larger uncertianties.
}
\label{simresults}
\end{center}
\end{figure*}

\subsection{Internal accuracy and precision} \label{simtests}

Fig. \ref{simresults} summarises our results from the test with the simulated PARSEC sample with ``high-res.-like'' observational errors, when using the following set of measured quantities as an input: $\vec{x} = \{\mh, \Teff, \log g, m, \mathrm{colors} \}$. In each panel except for the last, we show the relative deviations of our estimated values from the true distances, $(d_{\mathrm{BPG}}-d_{\mathrm{true}}) / d_{\mathrm{true}}$, as a function of the model parameters.

The first row of this Figure demonstrates that our distance errors are typically below 20\% and do not show any strong systematics with the measured quantities \Teff, [Z/H] and $\log g$ themselves. However, especially the upper right panel (distribution of distance errors in the \Teff$-\log g$ diagram) reveals that our distances for main-sequence and subgiant stars are much better recovered than for giants (mean deviation $-0.5\%$, root-mean-square scatter $10.6\%$ vs. $1.8\%$ and $23.3\%$, respectively). This is a known problem, because of the larger overlap of models with very different luminosities in the giant region of the Hertzsprung-Russell diagram (see Fig. \ref{simsample}).

The second row of Fig. \ref{simresults} examines the dependency of our distance errors with the model parameters which are not directly measured through spectroscopic observations: mass, age and distance. While there is no correlation between the distance errors and the true distances themselves, we do see some systematic trends with mass and age. In particular, our code tends to slightly overpredict the distances to very old, very-low-mass giant stars, while it underpredicts the distances for very young ($\tau < 100$ Myr) giants with super-solar masses. However, when taking into account the Galactic star-formation history and initial mass function (IMF), these groups typically represent a negligible minority of the stars targeted by large spectroscopic stellar surveys; their fraction is clearly overrepresented in our test sample. We can therefore be confident that our code delivers unbiased distance estimates for the vast majority of stars.

Finally, in the lower right panel of Fig. \ref{simresults}, we show the distance uncertainties for the simulated stars, again as a function of \Teff ~and $\log g$. Evidently, the internal precision of our method is much better for main-sequence stars than for giants (excluding the red-clump/horizontal-branch stars at $\log g \sim 2.5$, for which we also find accurate and precise distances). A comparison between the upper and lower right panels of this Figure also shows that our code determines reliable and unbiased statistical uncertainties.

\subsection{Sensitivity to input observables} \label{foursets}

\begin{figure}
\begin{center}
\includegraphics[width=0.5\textwidth]{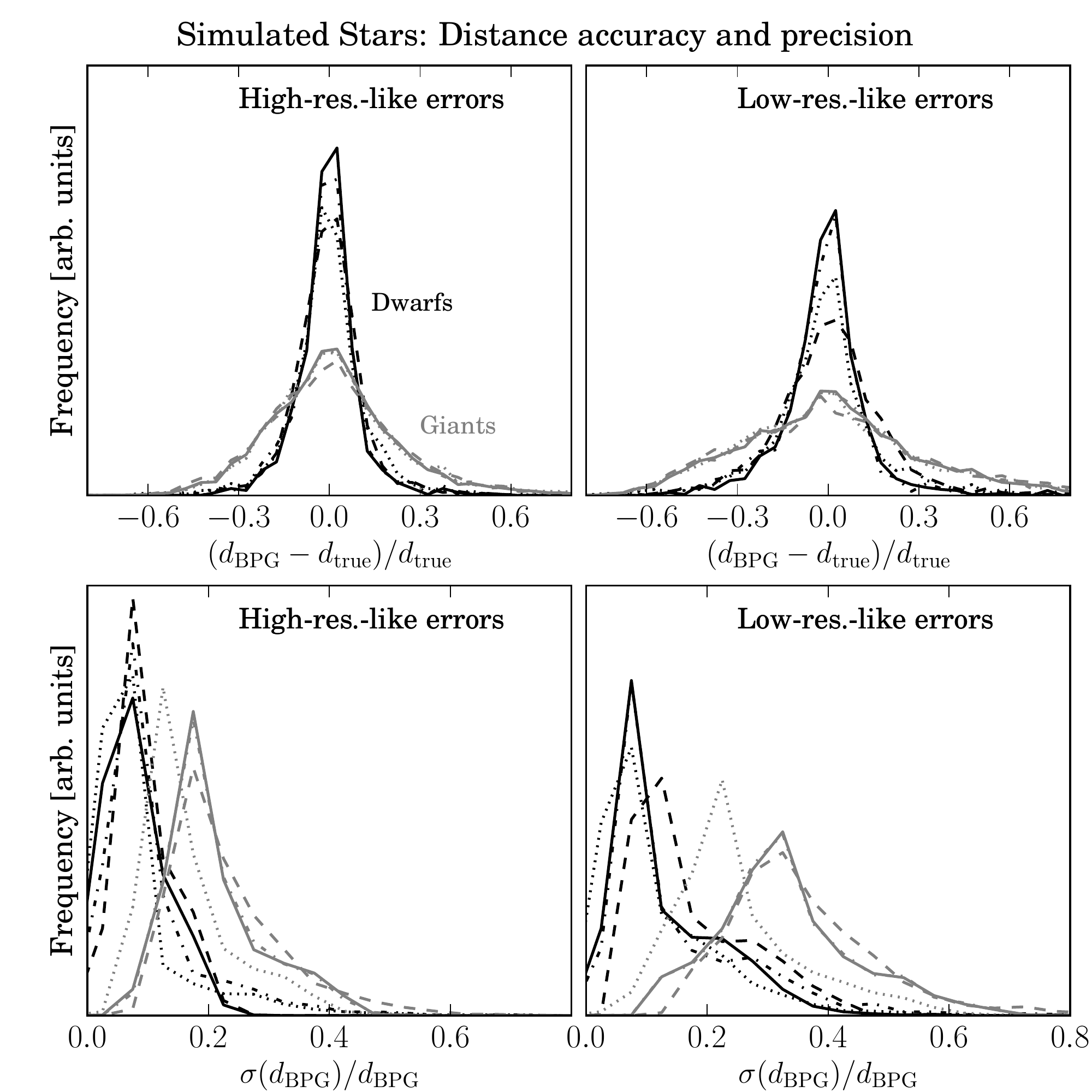}
\caption{\small Distance accuracy and precision for the four input observable sets introduced in Sec. \ref{foursets}. {\it Upper panels:} Distribution of relative distance errors, $(d_{\mathrm{BPG}}-d_{\mathrm{true}}) / d_{\mathrm{true}}$, for the sample of simulated stars drawn from PARSEC models (see Fig. \ref{simsample}) with ``high-resolution-like'' observational errors ({\it left}; see Sec. \ref{testset}) and ``low-resolution-like'' errors ({\it right}). In each panel, we are using the distances based on the four sets of measured parameters listed in Sec. \ref{foursets}. The results for dwarfs are shown in {\it black}, the results for the giants in {\it grey}. {\it Solid lines}: Set 1 ($\vec{x} = \{\mh, \Teff, \log g, m, \mathrm{colors} \}$); {\it dashed lines}: Set 2 ($\vec{x} = \{\mh, \Teff, \log g, m \}$); {\it dash-dotted lines}: Set 3 ($\vec{x} = \{\mh, \log g, m, \mathrm{colors} \}$); {\it dotted lines}: Set 4 ($\vec{x} = \{\mh, m, \mathrm{colors} \}$, with a cut at $|\Delta \log g| < 0.5$ dex).
{\it Bottom panels:} Distributions of our relative (internal) distance uncertainties, $\sigma(d_{\mathrm{BPG}})/d_{\mathrm{BPG}}$.}
\label{histos}
\end{center}
\end{figure}

\begin{figure*}
\sidecaption
\includegraphics[width=0.65\textwidth]{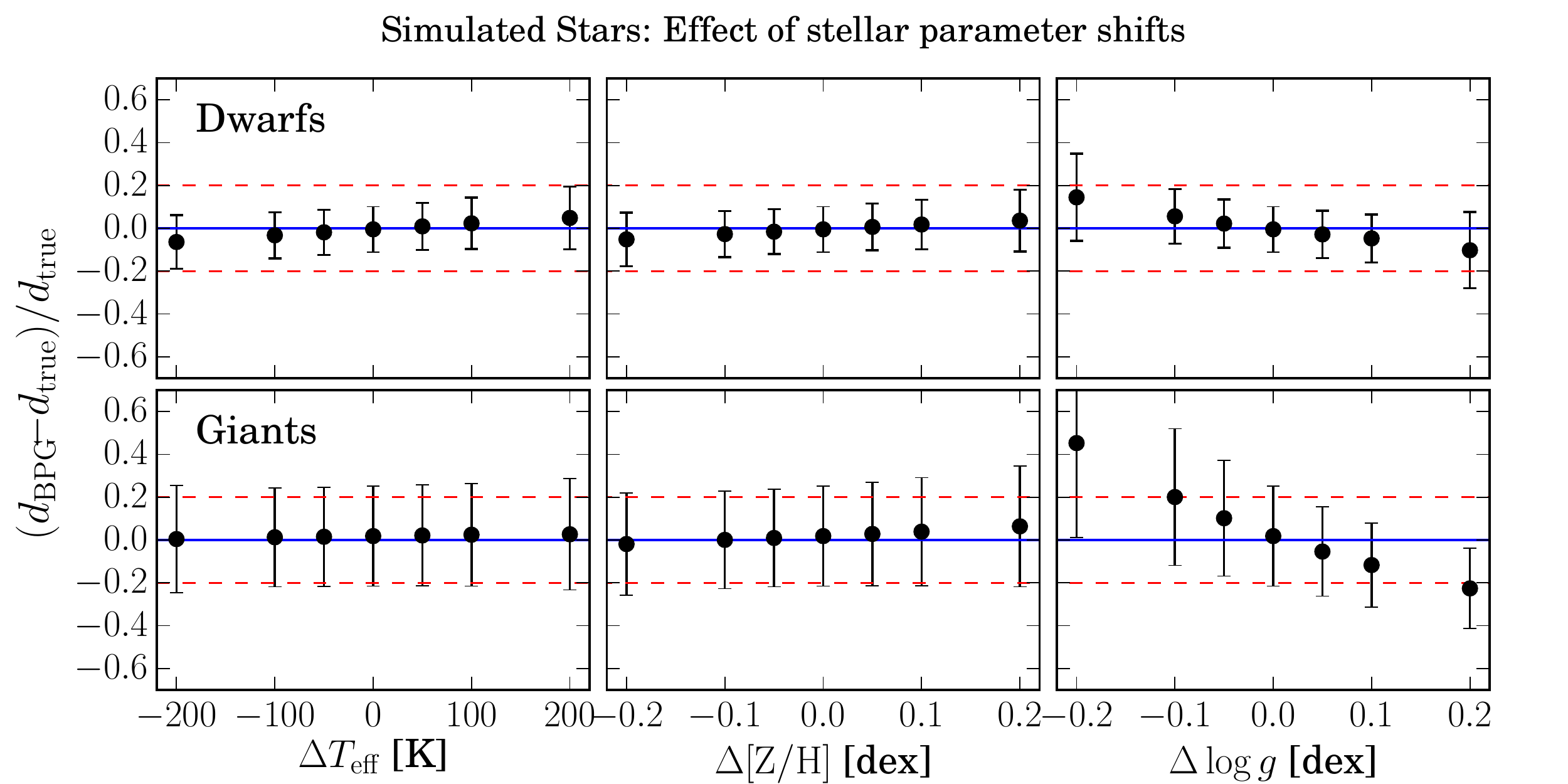}
 \caption{\small Effects of systematic shifts in the spectroscopically observed stellar parameters \Teff ~({\it left}), [Z/H] ({\it middle}) and $\log g$ ({\it right panels}) on our determined distances. For this exercise, we again use the simulated PARSEC sample with ``high-res.-like'' observational errors and the full input parameter set (Set 1: $\vec{x} = \{\mh, \Teff, \log g, m, \mathrm{colors} \}$). 
 In each panel, the errorbars show the mean and standard deviation of the relative distance errors, as a function of the value of a fixed shift in the particular stellar parameter.
 Dwarfs ($\log g>4$) and giants are displayed separately in the top and bottom rows, respectively.
}
\label{systematics}
\end{figure*}

Initial tests showed that, among the input measured quantities, $\log g$ is most critical to estimate unbiased distances,
since it is the quantity that best discriminates between low-luminosity dwarfs
and more luminous giants, which otherwise share similar values of temperature,
metallicities and colors. Dropping $\log g$ from the fitting procedure 
leads to overestimated distances for dwarf stars, and underestimated distances to giants, respectively. Similar tests also showed that cutting models with $\vert \vec{x_0} - \vec{x}| > 3 \vec{\sigma_x}$ decreases 
the processing time without appreciably changing the final distance estimates. 

Based on these initial assessments, we implemented a code that 
estimates distances using the method described above and based on 
four distinct sets $\vec{x}$:

\begin{enumerate}
\item $\vec{x} = \{\mh, \Teff, \log g, m, \mathrm{colors} \}$
\item $\vec{x} = \{\mh, \Teff, \log g, m \}$
\item $\vec{x} = \{\mh, \log g, m, \mathrm{colors} \}$
\item $\vec{x} = \{\mh, m, \mathrm{colors} \}$, but with a cut in models whose
$\log g$ differ by more than $0.5$ dex from the measured value.
\end{enumerate}

The code can accommodate other sets if so desired. The advantage of implementing different sets of parameters for distance determination is to provide flexibility: distances can be estimated even in the absence of one or two parameters from the set with the largest number of parameters (Set 1). Furthermore, the alternative distance estimates may be intercompared to provide a means to evaluate the sensitivity of the method to the particular combination of spectroscopic and photometric parameters adopted.

Often, the data set in a given sample includes only $\feh$ values, 
which are used in place of \mh. 
When $\afe$ values are available, we adopt the empirical relation $\mh = \feh + \afe$ for the comparison to PARSEC isochrones \citep[e.g.,][]{Anders2014}.
Another important issue is extinction: the photometric data,
$m$ and colors, must be corrected for extinction and redenning
before comparing with the models. The exact procedure depends on the bandpasses used, on the available extinction estimates and on the assumed extinction coefficients.

The top panels of Fig. \ref{histos} show the results of applying the method to our simulated observations (see Sec. \ref{testset}). The distribution of the relative distance errors, $(d_{\mathrm{BPG}}-d_{\mathrm{true}}) / d_{\mathrm{true}}$, are displayed for the four parameter sets listed above, for both the ``high-res.-like'' (left panel) and the ``low-res.-like'' simulations, and for giants and dwarfs separately. 

The four distance estimates proposed lead to reasonable results as attested by these error distributions. Surprisingly, there is only a very small systematic degradation in the precision of our distances as we move down from set 1 to 4. As noted above, set 1 in the ``high-resolution'' case yields mean distance errors of $-0.5\%$ with an rms scatter of $10.6\%$ for main-sequence stars ($\log g > 4.0$), while for giants the numbers are expectedly a bit worse ($+1.8\%$ bias and $23.3\% $ precision).

For the other three sets, these numbers are $-0.2\% \pm11.3\%$, $-0.7\% \pm13.1\%$, $-2.9\% \pm21.5\%$ for dwarfs, and $+1.7\% \pm24.3\%$, $+1.8 \pm23.4\%$, $+2.8\% \pm25.1\%$ for giants, respectively. This trend is expected, at least in the context of simulations, since usage of more measured parameters will tend to better constrain the stellar models that best describe each star. 

As can be seen in Fig. \ref{histos} (top right panel), these numbers are slightly worse for the ``low-resolution'' case. The relevant numbers for this and all other comparison samples are listed in Table \ref{summarytable}.
In the following, we will use only the full input parameter set (Set 1: $\vec{x} = \{\mh, \Teff, \log g, m, \mathrm{colors} \}$).

The bottom panels of Fig. \ref{histos} show the distributions of our distance uncertainties, $\sigma(d_{\mathrm{BPG}})/d_{\mathrm{BPG}}$, for each of the four input sets, and for dwarfs and giants separately. Again, we also show the numbers for the ``low-resolution'' case.
As expected, the uncertainty distributions for dwarfs peak at $<10\%$, while the distributions for giants are broader and peak at larger values.

\begin{figure*}
\centering
\includegraphics[width=0.79\textwidth]{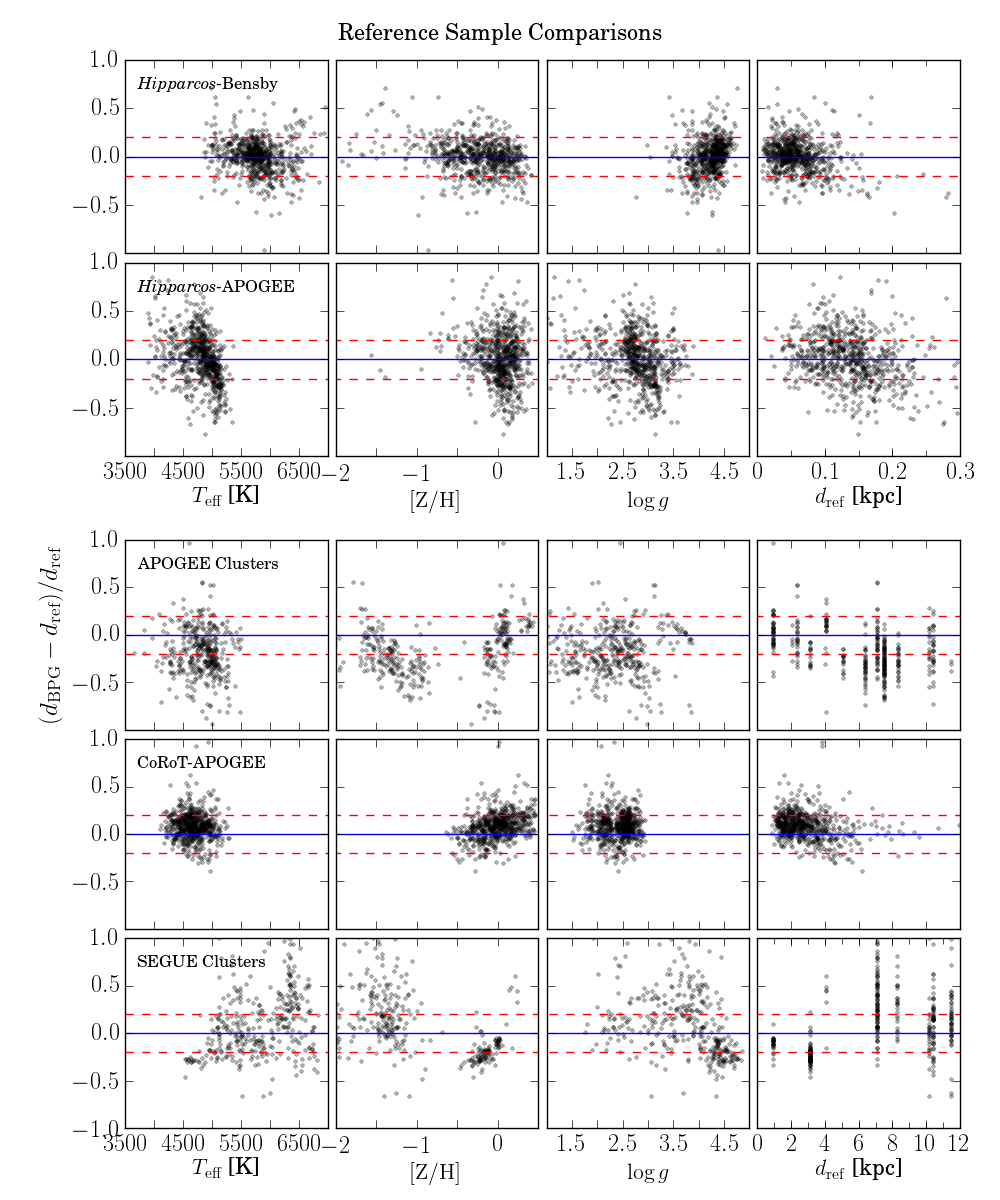}
\caption{\small Relative distance residuals (estimated -- reference) as a function of the main spectroscopic parameters \Teff, [Z/H] and $\log g$ as well as the reference distances $d_{\mathrm{ref}}$. 
{\it Top rows}: Comparison to the astrometric distance scale provided by the {\it Hipparcos} satellite, using the \citet{Bensby2014} FGK dwarf sample (1st row) and the APOGEE-{\it Hipparcos} giant sample (2nd row).
{\it 3rd row}: APOGEE clusters.
{\it 4th row}: CoRoT-APOGEE sample.
{\it 5th row}: SEGUE clusters. The central dashed line is the identity line, whereas the upper and lower ones indicate deviations at the 20\% level.}
\label{referencesamples}
\end{figure*}

\subsection{Effect of systematic stellar parameter errors} \label{sys}

In order to gain some insight into the robustness of our method to systematics in the stellar parameters used, we have studied the effect of constant shifts in each of the spectroscopic input parameters on our distance estimates. Fig. \ref{systematics} shows the results of this experiment: in each panel, 
all except for one input parameter are kept constant for the whole simulated  ``high-res.'' sample, while the remaining parameter is shifted by a certain amount, and the effect on the relative distance error (observed -- true) as a function of this shift. As before, we treat dwarfs and giants separately, because these two groups display distinct systematic trends.

The most striking observation from Fig. \ref{systematics} is that even a substantial systematic shift in the temperature ($|\Delta T_{\mathrm{eff}}|\lesssim 200$ K) or metallicity scale ($|\Delta {\mathrm{[Z/H]}}|\lesssim 0.2$ dex) does not terribly affect neither accuracy nor precision of our distances.
This is in stark contrast to systematic shifts in the surface gravity parameter (which are in fact quite commonly found even in high-resolution spectroscopic surveys -- e.g., \citealt{Holtzman2015}).

\subsection {The model priors} \label{priors}

In order to determine distances to stars in the Milky Way, we use our current knowledge about the Galaxy and its components
to build the model prior probability distribution, $P({\vec{x_0}})$.
This includes basic information concerning the
distributions of stars as a function of position, mass, age, and metallicity. 

In particular, we assume that all stars follow a Chabrier initial mass function \citep{Chabrier2003} and incorporate the probability that a randomly selected star falls within a given mass range, $p(m_{\star})$. This involves an integral over the mass function which is provided by the model isochrones. 

We take into account the effect of observing a given star in a small but finite solid angle, by applying a geometric prior $\propto s^2 \mathrm{d}s$ called the volume element, where $s$ is the model distance.
We further assume different age and spatial distributions for the basic
Galactic components, namely the thin disc, the thick disc, and the spheroid. We here adopt the same Galactic structural parameters as \cite{Binney2014a}, and refer to this work for a detailed justification. 

The age distributions, $p(\tau)$, are taken to be uniform with an upper
limit of 10 Gyrs for the thin disc, and lower limits of 8 and 10 Gyrs for the
thick disc and halo, respectively.

The spatial priors for the thin and thick discs are given, respectively, by
$$p_{\rm thin} (R,Z) \propto s^2 \mathrm{d}s \cdot \exp(-R/h_{R,\rm thin}) \cdot \exp(-\vert Z \vert / h_{Z,\rm thin}),$$
$$p_{\rm thick} (R,Z) \propto s^2 \mathrm{d}s \cdot \exp(-R/h_{R, \rm thick}) \cdot \exp(-\vert Z \vert / h_{Z,\rm thick}), $$

\noindent where $(R,Z)$ are cylindrical coordinates with origin at the Galactic center, whose values are computed given the model distance $s$ and a direction $(l,b)$, and $h_R$ and $h_Z$ are the respective scale lenghts and heights. In computing the cylindrical coordinates, we use $R_0 = 8.33$ kpc as the distance from the Sun to the Galactic Centre. 

For the halo, we assume a power-law density profile with spherical symmetry:

$$p_{\rm halo} (r,l,b) \propto s^2 \mathrm{d}s \cdot r^{-\gamma}, $$

\noindent where $r(s,l,b)$ is a radial spherical coordinate. Again, as in
\cite{Binney2014a}, we adopt $\gamma =$ 3.39.

The three density profiles are normalised at the solar location, which,
for simplicity, we take to be $(R,z)=(R_0,0)$ and $r=R_0$, respectively,
in cylindrical and spherical coordinates. The normalization values are
again from \cite{Binney2014a}. 

We assume that the three basic Galactic components follow Gaussian metallicity
distribution functions (MDFs), $p_i({\mathrm{[Z/H]}})$, exactly as described by \cite{Binney2014}. Our final model prior is then given as

$$P({\vec{x_0}}) = p(m_{\star}) \sum_i p_i(\tau)~p_i(\vec{r})~p_i([Z/H]).$$

\section{External Validation} \label{validation}

\begin{table*}
\centering
\caption{Summary of the reference data: parameter ranges, uncertainties, and provenance. $\log g$ values in the CoRoT-APOGEE sample are based on asteroseismic measurements. All other non-photometric parameters were derived spectroscopically; the relevant references are given in the footnotes indicated in the third column.}
\tiny
\setlength{\tabcolsep}{2pt}
\begin{tabular}{cccccccccc}
\hline
\hline
Sample & dist. range [kpc]  & $\sigma(\Teff)$ [K] & $\Teff$ range [K] & $\sigma \log g$ & $\log g$ range & $\sigma$ [Fe/H] & [Fe/H] range & photom. & magnitude range \\
\hline 
Simulated sample & $0.1-6$ & $100$ & $3000-7000$ & 0.1 & $1.0-5.0$ & 0.1 & $-2.2-0.5$ & $JHK_s$ & distance-limited \\
{\it Hipparcos}-Bensby & $<0.3\,^1$ & $30-100\,^2$ & $4800-7000$ & 0.1 & $3.0-4.8$ & 0.1 & $-2.5-0.5$ & $B_TV_T\,^3$ & $V<9\,^2$ \\
{\it Hipparcos}-APOGEE & $<0.3\,^1$ & $90$ & $3800-5500$ & 0.2 & $1.0-4.0$ & 0.1 & $\sim-1.0-0.5$ & $JHK_s\,^5$ & $(J-K)_0>0.5;H<7$ \\
CoRoT-APOGEE & $1-6\,^4$ & $100\,^4$ & $4000-5300$ & 0.05 & $1.6-3.0$ & 0.1 & $-0.8-0.5$ & $JHK_s\,^5$ & $(J-K)_0>0.5;H<13.8\,^4$ \\ 
APOGEE Clusters & $1-6\,^{6,7}$ & $100\,^8$ & $3700-5000$ & 0.2 & $1.0-3.4$ & 0.1 & $-1.0-0.5$ & $JHK_s$ & $(J-K)_0>0.5;H<13.8\,^9$ \\
SEGUE Clusters & $1-12\,^{6,7}$ & $130\,^{10}$ & $4700-7000$  & 0.21 & $1.6-5.0$ & 0.11 & $-3.0-0.0$ & {\it ugriz} $^{11}$ & $0.48<(g-r)_0<0.55;r_0<20.2\,^{12}$ \\
\hline
\end{tabular}
\tablebib{
\tiny
(1) \citet{vanLeeuwen2007}; 
(2) \citet{Bensby2014};
(3) \citet{Hog2000a};
(4) Anders et al. (2015) and references therein;
(5) \citet{Cutri2003};
(6) WEBDA Database;
(7) \citet{Harris1996};
(8) \citet{Ahn2014};
(9) \citet{Zasowski2013};
(10) \citet{Ahn2012};
(11) \citet{Fukugita1996};
(12) \citet{Lee2008}
}
\label{sampletable}
\end{table*}

In this section, we present results of the application of our method 
described in the previous section to a number of reference samples 
for distance determination. These include distances computed from astrometric parallaxes obtained by ESA's {\it Hipparcos} mission \citep{Perryman1989, Perryman1997, vanLeeuwen2007}, precise asteroseismic distances produced by the CoRoT-APOGEE collaboration for a set of solar-like oscillating red giant stars (Anders et al. 2015, subm.), and distances to well-studied open and globular clusters. 
As in the previous sections, we use $d_{\mathrm{BPG}}$ to refer to our distances in all figures. Those taken as reference are referred to as $d_{\mathrm{ref}}$. In Table \ref{sampletable}, we provide information about the sources, derivation methods and precision of the parameters in the reference samples. The results of the comparison with the reference samples are summarised in Fig. \ref{referencesamples} and Table \ref{summarytable}. We also compare our distance estimates with those obtained by \cite{Binney2014a} for a sample of giant stars from the RAVE survey. This last comparison represents a cross-check of the implementation of the algorithm rather than a benchmark test, as the authors follow a similar theoretical background. 

Fig. \ref{referencesamples} presents the comparison of our spectro-photometric distances (BPG) to those from the reference samples. The first three panels of each row show our relative distance residuals with respect to the reference set, as a function of the main spectroscopic parameters \Teff, [Z/H] and $\log g$. The last panel of each shows the residuals as a function of  the reference distance. Each row corresponds to a particular reference sample discussed in this Section. 

\subsection {Comparison with the Hipparcos scale - The FGK star sample of \citet{Bensby2014}}
\label{hiparcoscomp}

As a first validation test of our spectro-photometric distance algorithm, we use the high-resolution FGK dwarf sample of 714 solar-neighborhood stars with well-determined atmospheric parameters by \citet{Bensby2014}. Hereafter, we will refer to this sample as the {\it Hipparcos}--Bensby sample, because for these stars precision parallaxes from {\it Hipparcos} \citep{vanLeeuwen2007} are available. We use the optical photometry from the Tycho-2 catalogue \citep{Hog2000a}, and set $mag = V_T$, $\Mabs = M_{V_T}$ and colors = $\{B_T-V_T\}$ in the set of parameters to be compared to the models. 
As throughout this paper, the models used for this comparison are from the PAdova TRieste Stellar Evolution Code \citep[PARSEC 1.2S,][]{Bressan2012, Tang2014, Chen2015}, with a narrow metallicity step of $\mh = 0.1$ and age steps of $\log \tau (yrs) = 0.05$. 

The top row of Fig. \ref{referencesamples} shows the results of our distance code for the {\it Hipparcos}--Bensby sample. There is little or no systematics in the residuals with the parameters themselves. The mean and rms residuals over all stars are $0.4\%$ and $18.2\%$. In fact, the observed rms residual is comparable to the expected errors in the parallaxes and in our distances based on the ``high-res.'' simulation (presented in Sec. \ref{testset}) combined together. Because the {\it Hipparcos}--Bensby sample combines high-resolution spectroscopic data for nearby dwarf stars with accurate parallaxes, it is arguably the best reference sample to test our code against.

We have validated that, as in our simulation tests described in Sec. 2, the four sets of parameters yield acceptable distances. In the
subsequent validation analyses in this paper, we will concentrate on the full parameter set $\{$\Teff, [Z/H], $\log g, \mathrm{colors}\}$, since it makes full use of the set of spectroscopic and photometric parameters which is common to the surveys to which we are applying our method.

\subsection {APOGEE}
\label{apogeecomp}

The SDSS-III project APOGEE has acquired high-resolution near infra-red (NIR) spectra ($R \simeq 22,500$) of $> 100,000$ stars selected from the 2MASS Point Source Catalogue \citep{Cutri2003}, most of which are
red giant stars located at low Galactic latitudes. APOGEE is providing
a unique spectroscopic sample of disc-dominated stars with unprecedented 
volume coverage, for which precision kinematical and abundance 
measurements are available. For more details about APOGEE data, we refer to \citet{Ahn2014, Alam2015}, and \citet{Holtzman2015}.
Validation of APOGEE stellar parameters partly relies on calibrating star clusters with many probable member stars with well-determined parameters \citep{Meszaros2013, Holtzman2015}. 
In this section we use two APOGEE subsamples to test our distances.
In Sec. \ref{corotcomp}, we use a set of 678 CoRoT stars in the exoplanet fields 
LRa01 and LRc01 \citep[e.g.,][]{Miglio2013} which have been co-observed within an APOGEE ancillary campaign, and whose asteroseismic analysis has provided high-precision surface gravities (Anders et al. 2015, subm.). In
Sec. \ref{apogeecluscomp}, we make use of the APOGEE calibration cluster dataset. 

In the case of APOGEE, we set $m = K_s$,
$\Mabs = M_{Ks}$, and colors $= \{ (J-H), (H-K_s) \}$, all of which are
corrected for extinction before computing the distances. For the cluster stars, extinction values are based on the Rayleigh-Jeans Color Excess method \citep[RJCE,][]{Majewski2011, Nidever2012a, Zasowski2013}. This method is based on stellar colour excesses measured using fluxes in near and mid infra-red bands, where almost all stellar spectra are in the Rayleigh-Jeans regime and therefore have very similar intrinsic colours. For the stars in common with CoRoT, we use the precise extinctions calculated by Anders et al. (2015), using the method described in \cite{Rodrigues2014}.

We use the calibrated stellar parameters determined by the APOGEE Stellar Parameters and Chemical Abundances Pipeline (ASPCAP; \citealt{Holtzman2015}, Garc\'ia P\'erez et al. 2015, subm.). Some minor improvements on the DR12 calibrations of metallicity and surface gravity have been applied; they are described in Appendix \ref{apoapp}. Since \afe~is again available, we assume that $\mh = \feh + \afe$ for APOGEE stars\footnote{We here refer to ASPCAP's cluster-calibrated \mh$~$ values \citep{Alam2015, Holtzman2015} as \feh, as ASPCAP metallicities were calibrated on literature \feh values.}

For the comparison with the CoRoT sample (Sec. \ref{corotcomp}), APOGEE surface gravities were replaced by those from asteroseismic scaling relations \citep[e.g.,][]{Miglio2013}. Table \ref{sampletable} summarises the data provenance.

\begin{table*}
\caption{Summary of the results for the reference distance samples used in this work.}
\setlength{\tabcolsep}{9pt}
\begin{footnotesize}
\begin{tabular}{lcccc}
\hline\hline\\
\smallskip
Sample & {$N_{\mathrm{stars}}$} & {Dist. Range [kpc]} & {Mean rel. residuals [\%]} & {RMS rel. residuals [\%]} \\
\hline
  Simulated ``high-res.'' stars & 5000 & $0.1-6$ & $+1.2$ & $20.9$ \\
  \qquad Dwarfs ($\log g>4$) & 1248 & $0.1-6$ & $-0.5$ & $10.6$ \\
  \qquad Giants ($\log g<4$) & 3752 & $0.1-6$ & $+1.8$ & $23.3$ \\
  Simulated ``low-res.'' stars & 5000 & $0.1-6$ & $+3.3$ & $33.4$ \\
  \qquad Dwarfs ($\log g>4$) & 1248 & $0.1-6$ & $-0.6$ & $15.5$ \\
  \qquad Giants ($\log g<4$) & 3752 & $0.1-6$ & $+4.6$ & $37.4$ \\
  {\it Hipparcos}-Bensby dwarfs & 714 & $\lesssim 0.2$ & $+0.4$ & $18.2$ \\
  {\it Hipparcos}-APOGEE giants & 712 & $<0.3$ & $+1.6$ & $26.4$ \\
  CoRoT-APOGEE giants & 644 & $1-12$ & $+8.7$ & $14.9$ \\
  APOGEE clusters -- cluster age prior & 392 & $1-15$ & $-16.5$ & $29.9$ \\
  SEGUE clusters -- field priors & 419 & $1-12$ & $+14$ & $43$ \\
  \qquad $\Delta$ log g =+0.25 dex; field priors & 425 & $1-12$ & $-2$ & $35$ \\
  \qquad $\Delta$ log g =+0.25 dex; clus. age priors & 385 & $1-12$ & $+13$ & $39$ \\
\hline
\end{tabular}
\end{footnotesize}
\label{summarytable}
\end{table*}

\subsubsection{The Hipparcos--APOGEE sample} \label{apohip}

Through a fiber feed from the New Mexico State University 1m telescope at Apache Point Observatory to the APOGEE spectrograph, it is also possible to use APOGEE to observe smaller numbers of bright stars, one at a time \citep[][Sec. 2.3]{Holtzman2015}. 
Apart from calibration, this setup is also being used within an ancillary campaign to observe several hundred red giant stars with measured {\it Hipparcos} parallaxes (Feuillet et al, {\it in prep.}). 
The data processing and stellar parameter determination works in exactly the same way as for the data taken with the SDSS 2.5m telescope.

We have used the DR12 {\it Hipparcos}--APOGEE sample of 712 stars with precise parallaxes ($\sigma(\pi)/\pi<10\%$) to fundamentally validate our stellar distance estimates also for giant stars. The second row of Fig. \ref{referencesamples} shows our results. The mean and rms residuals amount to $+1.6\%$ and $26.4\%$, respectively.

Recalling the tests performed in Sec. \ref{method}, we find that these numbers are very much comparable with the trends found for the giant sample of simulated ``high-resolution'' stars (see Table \ref{summarytable}). Taking further into account that the {\it Hipparcos} parallaxes for these relatively distant objects are also affected by uncertainties of order $10\%$, the results indeed supersede  expectations. 
Furthermore, we find no systematic trends with either of the stellar parameters.

\subsubsection{APOGEE - comparison with asteroseismology: the CoRoT-APOGEE sample} \label{corotcomp}

Another important test for our distance method is the comparison with asteroseismically derived distances. 
It has been shown by recent studies \citep[e.g.,][]{Miglio2012, SilvaAguirre2012, SilvaAguirre2013} that stellar distances
(or equivalently, radii) determined from CoRoT and Kepler lightcurves, either via asteroseismic scaling relations or by comparing asteroseismic parameters to predicted values from a grid of models, agree within a few percent with Hipparcos parallaxes and eclipsing binary data. For example, \citet{SilvaAguirre2012} show that distances can be derived with 5\% precision for solar-like stars by coupling the infra-red flux method to determine \Teff$~$ and bolometric fluxes to the grid-modelling of the measured values of the frequency of maximum oscillation power, $\nu_{\mathrm{max}}$, and of the frequency difference between dominant oscillation modes, $\Delta\nu$.
We therefore compared our spectro-photometric distances based on APOGEE stellar parameters with the distances obtained from CoRoT data for 678 stars in the fields LRa01 and LRc01, which have been successfully observed by both instruments. The reference distance scale is that of Anders et al. (2015) who used the well-tested stellar parameter estimation code PARAM \citep{daSilva2006, Rodrigues2014}, which is based on a similar theoretical background as this paper, but differs in a number of details.
PARAM utilises the parameter set $\{\Teff, \mathrm{[Z/H]}, \Delta\nu, \nu_{\mathrm{max}} \}$ to first estimate the posterior probability over the stellar models (i.e., it delivers mass, radius, age and absolute magnitude pdfs). In a second separate step, it compares the derived absolute magnitude pdfs with a number of broad-band photometric measurements to derive individual stellar distances and extinctions. The code has been extensively tested in \citet{Rodrigues2014} and Anders et al. (2015); it delivers very precise distances for asteroseismic targets ($\sigma(d)\lesssim2\%$).  
Because we use the same data and similar methods to derive stellar distances, the distance scale of Anders et al. (2015) does not provide a completely independent benchmark. However, it serves as an important cross-check for the implementation of our code.

The comparison is shown in the 4th row of Fig.~\ref{referencesamples}. We again show the relative distances residuals between our distances (BPG) and those from CoRoT-APOGEE. The general picture is reassuring, despite the limited number 
statistics. Our code converged for $644$ stars, the mean and rms relative distance residuals are $+8.7\%$ and $14.9\%$, respectively. No strong trends are seen with metallicity, effective temperature or surface gravity. 

The systematic differences between the two methods may arise from the different handling of priors, while the rms scatter is more likely to be attributed to our use of only three photometric passbands instead of the multi-wavelength photometry used by PARAM, and to minor differences in building the distance pdf. For simplicity, and because of the high quality of acquired data, Anders et al. (2015) opted to use an IMF prior plus flat priors on Galactic structure, to derive distances. However, their sample extends over a large range of Galactocentric distances, so that small systematic shifts like the one measured here are not excluded.

In summary, we find a very good agreement between the two methods, modulo a small systematic trend which we attribute to our choice of more comprehensive priors, and a scatter of $\sim15\%$ which could be reduced by the inclusion of more photometric observations in our dataset.

\begin{table}
\caption{Summary of the properties of star clusters used as reference for APOGEE and SEGUE distance estimates: dereddened cluster distance moduli, ages, metallicities, reddenings, and numbers of stars observed by APOGEE and SEGUE. Almost all of the listed clusters were used for APOGEE calibration, while SEGUE targeted a larger number of cluster members in some selected calibration clusters.}
\setlength{\tabcolsep}{5pt}
\begin{footnotesize}
\centering 
\begin{tabular}{lllllll}
\hline
\hline \\
  \multicolumn{1}{l}{Cluster} &
  \multicolumn{1}{c}{$(m-M)_0$} &
  \multicolumn{1}{c}{$\log(\tau)$} &
  \multicolumn{1}{c}{\feh} & 
  \multicolumn{1}{c}{E(B-V)} &
  \multicolumn{1}{c}{$N_{_{\mathrm{APO}}}$} & 
  \multicolumn{1}{c}{$N_{_{\mathrm{SEGUE}}}$} \\[.2pt]
\hline \\
NGC 188  & 11.55 & 9.63 & $-$0.02 & 0.082 &  5 &     \\
NGC 2158 & 13.52 & 9.02 & $-$0.23 & 0.36  & 15 &  10 \\
NGC 2420 & 12.45 & 9.05 & $-$0.26 & 0.029 & 16 & 125 \\
NGC 5466 & 16.02 &10.13 & $-1.98$ & 0.0   &  5 &     \\
NGC 6791 & 13.06 & 9.64 & $+0.15$ & 0.117 & 22 &  37 \\
NGC 6819 & 11.86 & 9.17 & $+0.07$ & 0.238 & 23 &     \\
NGC 7789 & 11.84 & 9.23 & $-$0.08 & 0.217 &  4 &     \\
M 2      & 15.50 &10.11 & $-1.65$ & 0.06  &  3 &  64 \\
M 3      & 15.07 & 9.69 & $-1.5$  & 0.12  & 18 &  35 \\
M 5      & 14.46 &10.02 & $-1.29$ & 0.03  & 84 &     \\
M 13     & 14.33 &10.06 & $-1.53$ & 0.02  & 46 & 149 \\
M 15     & 15.39 &10.07 & $-2.37$ & 0.10  & 31 &  78 \\
M 35     & 10.37 & 7.97 & $-0.16$ & 0.262 &    &   6 \\
M 53     & 16.32 &10.10 & $-2.1$  & 0.02  & 13 &     \\
M 67     &  9.79 & 9.41 &  0.00   & 0.059 & 48 &  50 \\
M 71     & 13.01 &10.00 & $-0.78$ & 0.25  &  2 &     \\
M 92     & 14.65 &10.14 & $-2.31$ & 0.02  & 34 &  39 \\
M 107    & 15.05 &10.14 & $-1.02$ & 0.33  & 39 &     \\
Be 29    & 15.86 &9.025 & $-0.44$ & 0.157 &  9 &     \\
  \hline
\end{tabular}
\end{footnotesize}
\label{clustable}
\end{table}

\subsubsection{APOGEE - Cluster comparison}
\label{apogeecluscomp}

In the third row of Fig. \ref{referencesamples}, we compare our spectro-photometric distances with those
obtained from isochrone fitting of star cluster 
color-magnitude diagrams (CMDs). We again restrict the comparison to the spectro-photometric distances based on the parameter set (1) in 
Sec. \ref{method}. We use a subsample of the $\sim400$ 
open and globular cluster stars that are 
used for calibration of ASPCAP, as described in \citet{Meszaros2013, Holtzman2015}. We again refer to Table \ref{sampletable} for a summary of the data used. Most of the stars with reliable parameters belong 
to the globular clusters M5, M13, M15, M92, M107, and to the open cluster M 67. The cluster distances and ages are adopted from the WEBDA cluster database in the case of open clusters and \cite{Harris1996} for the globular clusters. Table \ref{clustable} summarises the cluster properties.

The right panel in this comparison shows a hint that our distances are being underestimated for more distant clusters. The global mean residual is $-16.5\%$, with an rms scatter of $30\%$. All clusters, except for NGC 2420 ($d_{\mathrm{ref}} \simeq 3$ kpc), have most of their stars with residuals smaller than 30\%. 
Instead of the field star priors discussed in Sec. \ref{priors}, we have used a simple age prior for each cluster. This adopted cluster prior simply assumes a lognormal distribution of cluster stellar ages, whose mean is the literature age estimate for the cluster and the dispersion is taken as $d\log \tau$ = 0.1. We find that for most APOGEE clusters this prior is irrelevant, but the age prior improves the agreement with the reference distances for NGC 2420 and NGC 2158, while it negatively affects M71, which is the oldest and most metal poor cluster in the list.

The comparison suggests that our spectro-photometric method, despite the 
scatter, yields distances that are in general agreement with the cluster scale, but subject to significant systematics, especially as a function metallicity (see Fig. \ref{referencesamples}, third row, second panel). The fact that we do not see these systematic trends in the comparisons with the {\it Hipparcos} scale suggests that the cluster distances might be subject to improvements in the underlying stellar physics (e.g., the inclusion of rotation; \citealt{Brandt2015}), and less of a gold standard than commonly assumed. Another important caveat is the possible contamination of the cluster sample with non-members. 

Regarding the uncertainty of isochrone-based cluster distances, it has been shown by \citet{Pinsonneault2000} that open cluster 
distances (which are usually determined via measuring the main-sequence shift relative to the Hyades cluster) may be 
subject to zero-point shifts, based on changes in the adopted distance to the Hyades and interstellar reddening, as well 
as to the metal and helium abundance, and the age, of the Hyades \citep[e.g.][]{Brandt2015b}. However, assuming a conservative error of 0.1 mag for the cluster 
distance moduli, the uncertainties in the spectroscopic distances are still by far the most important.

\begin{figure}
\includegraphics[width=0.5\textwidth]{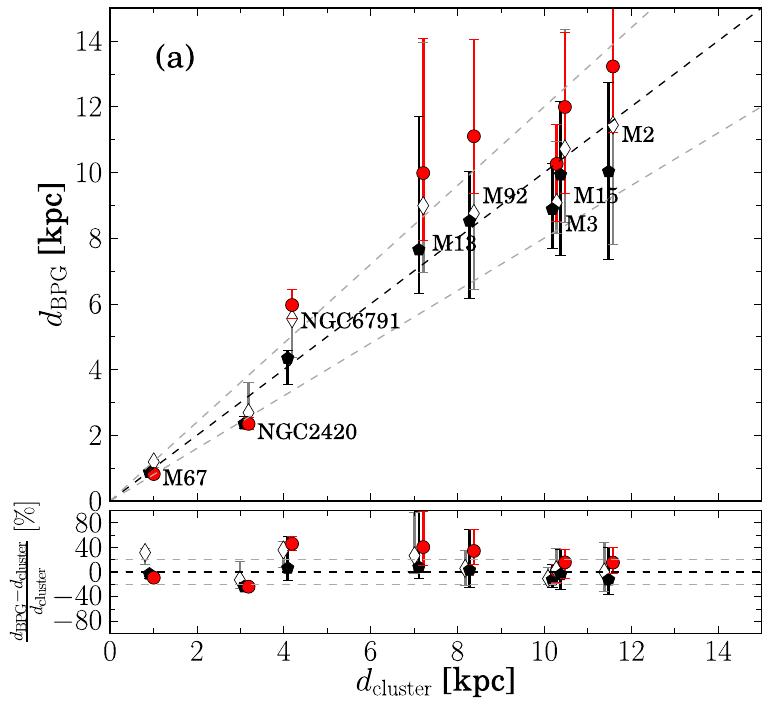}
\caption{Comparison of our distances and the host cluster distances $d_{\mathrm{cluster}}$ for a subsample of clusters with SEGUE observations. For each cluster the median distance and the 68\% quantiles (as error bars) are shown assuming the following different scenarios: {\it open diamonds} - uncalibrated SSPP DR9 parameters with field priors,
{\it black pentagons} - a $\log g$ shift of 0.25 dex and field priors, and 
{\it red circles} - a shifted $\log g$ and a simple cluster age prior. For visibility the different symbols are slightly offset with respect to $d_{\mathrm{cluster}}$. The lower panel shows the distance residuals, similarly to the right column of panels in Fig. \ref{referencesamples}.
}
\label{segueclus}
\end{figure}

\subsection{SEGUE Clusters} \label{seguecomp}

The Sloan Extension for Galactic Understanding and Exploration (SEGUE) is
a large optical spectroscopic survey at low resolution \citep[$R \simeq 2,000$,][]{Smee2013}. 
Its goal is to deepen our knowledge about the Galactic structural
components and their stellar content, 
sampling them mostly at high latitudes \citep{Yanny2009, Eisenstein2011}. In this sense,
SEGUE is largely complementary to APOGEE in terms of coverage of the Galactic
components. For more info on the spectra and instruments we refer to \cite{Gunn2006} and \cite{Smee2013}.
SEGUE data have been processed through the SEGUE Stellar 
Parameter Pipeline (SSPP), which is described, along with its
improvements, in a series of papers \citep{Lee2008,Lee2008a,AllendePrieto2008a,Smolinski2011,Lee2011a}. 
Particularly important to the SSPP is the 
validation of the derived stellar parameters.
Field stars with high-resolution spectra and 
known members of well studied star clusters have been used
for that purpose \citep{AllendePrieto2008a,Lee2008a,Smolinski2011}. 

As in the case of APOGEE (see Sec. \ref{apogeecluscomp}), we here use 
the sample of cluster stars, whose distances are well-known from
isochrone fitting, to further
test our distance estimates. 
In the case of SEGUE data, we set mag $= g$,
$\Mabs = M_g$, and colors $= \{ (u-g), (g-r), (r-i), (i-z) \}$ (see \cite{Gunn1998} for a description of the SDSS camera and \cite{Fukugita1996} for information on the photometric system).
The spectroscopic and photometric data are from the SDSS 
{\it Ninth Data Release} \citep[DR9;][]{Ahn2012} database. 
Since $\afe$ is available for all SEGUE stars in our reference clusters, 
we assume that $\mh = \feh + \afe$ in order to compare the data to
the model $\mh$ values.
As previously mentioned, all measured quantities have associated uncertainties.
The measured photometric quantities also must be corrected for 
extinction and reddening to allow a direct comparison to the models.
Again, the data for the clusters used here are listed in Table \ref{clustable}.

The original sample contained 11 clusters, among open and globular, totalling a bit over 1000 stars. For 593 stars, a complete set of spectroscopic and photometric parameters and associated uncertainties allowed us the use of parameter set (1) to estimate distances. As the sample is spectroscopic, in all other cases the missing parameter(s) was(were) photometric. Distances were successfully computed (i.e., at least one model was found within $3~\sigma$ of all the measured quantities) for 425 out of the 593 stars, pertaining to the eight clusters listed in the table. We include only those clusters with at least five confirmed members and successfully derived distances using our approach. 
Their distances, ages and metallicities were again taken from
the WEBDA database in the case of open clusters, and from \cite{Harris1996}
in the case of the globular clusters. 

\begin{figure*}
\begin{center}
\includegraphics[width=0.9\textwidth]{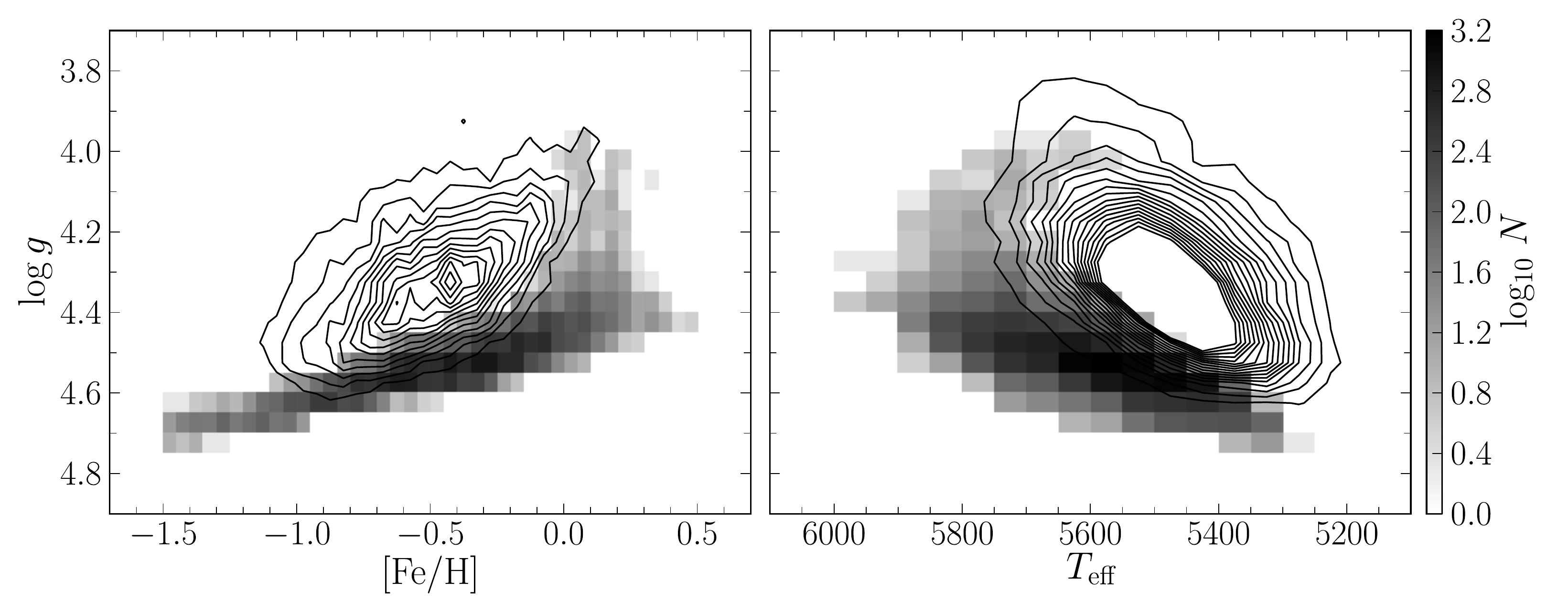}
 \end{center}
\caption{Left panel: The contours show the density of SEGUE dwarfs in the 
$\log g~vs.~$ \feh$~$ plane. The colour density map corresponds to the same type 
of stars simulated by the TRILEGAL code. Right panel: same as in the previous
panel, but stars are shown in the $\log g~vs.~\Teff~$ plane.}
\label{loggshift}
\end{figure*}

In the last row of Fig. \ref{referencesamples}, we show the distance residuals of individual stars against \feh, \Teff, $\log g$, and reference cluster difference, similarly to what we did with the other samples in this Section. The scatter is larger than in previous samples, attesting the lower resolution of the data. Still, most of the stars have residuals close to or within the 20\% lines and do not exibit any strong systematics with the spectroscopic parameters. The more deviant points correspond to a fraction of the low metallicity (\feh $~< -1.0$) giant and subgiant ($\log g \leq 4.0$) stars, located beyond $\simeq$ 6 kpc.
The mean and rms residuals over all stars are $0.14 \pm0.43$ after applying a 3-$\sigma$ clipping. Our individual distances are biased towards large values for a significant fraction of the stars in the more distant globular clusters. This trend is opposite to what was observed in the case of APOGEE clusters (see Sec. \ref{apogeecluscomp}). If we consider only the 3 clusters with $d_{\mathrm{ref}} < 5$ kpc, the mean and rms scatter are reduced to  $0.03 \pm0.27$. 

We investigated further the cause of this systematic overestimate
of the distances to part of the SEGUE stars in the more distant clusters. In analyzing the distribution
of different SEGUE spectroscopic parameters, it became clear that the $\log g$ parameter for our sample, as published in DR9, appears to be underestimated for most stars in those clusters. 
Recalling that $\log g$ is the most important quantity to efficiently
separate dwarf and giant stellar models, we have decided to heuristically correct for this shift in this work. 

The contours (black) in Fig. \ref{loggshift} show the distribution of a large sample of SEGUE G-dwarf stars 
(sample compiled from DR9 and described and analysed in Brauer 2015 (PhD thesis, subm.)) in the 
$\log g~ vs.~ \Teff$ (left panel) and $\log g~ vs.~ \feh$ (right panel) plane. For comparison, 
we also present the density distributions for a mock sample of G-dwarf stars obtained from 
TRILEGAL simulations \citep{Girardi2005, Girardi2012} covering the same SEGUE plates as present in the SEGUE G-dwarf data sample. These simulations were carried out using ADDSTAR, a web based tool that uses parallel processing to efficiently run TRILEGAL in many independent pointings \citep{Balbinot2012}.
The TRILEGAL G-dwarf mock sample shown here has been compiled by applying the same overall cuts to the simulation that were used for the SEGUE G-dwarf data sample including the colour-magnitude limits for SEGUE G-dwarfs. For a detailed description of the TRILEGAL mock sample see Brauer (2015). 

A clear shift is seen in both panels, in the sense that
the data have systematically lower values of $\log g$ for a fixed 
\Teff$~$ or metallicity by $\simeq 0.25$ dex. The comparion with the population synthesis model reveals that the SSPP log $g$ values published in DR9 are underestimated, and that shifting the $\log g$ values for the SEGUE stars
is clearly needed, at least for the purpose of deriving reliable distances. Further details about the TRILEGAL mock G-dwarf sample, the discrepancies 
seen in stellar parameters between data and simulation, and the underestimated surface gravity values published with DR9 are given in Brauer (2015).

We thus decided to test the
effect of applying a shift of 0.25 dex to the $\log g$ values when computing the
distances. The results are shown in Fig. \ref{segueclus}, where we
compare the median spectro-photometric distances of each cluster with the reference cluster distances for different $\log g$ calibrations and/or priors used. The open diamonds correspond to median cluster distances using the SEGUE data as they come from DR9. The black pentagons show the effect of shifting $\log g$ as explained above. The global mean offset relative to the reference distances has been reduced to $-2.4\%$, although the rms relative residual remains high (35\%). For the three more nearby clusters, the distances become systematically underestimated by $-18\%$ although the scatter around the mean residual has decreased considerably to $12\%$. The result of adopting a simple age prior for the clusters, as done for APOGEE data, is shown as the red circles of Fig. \ref{segueclus}. This choice has a sizable effect on the distances in the sense of increasing them for the more distant clusters, yielding a mean and rms relative residual of $+0.13 \pm0.39$. 

We conclude that, despite the large scatter in distances of individual cluster stars, there is a general agreement between our median distances for SEGUE cluster stars and those from CMD fitting. A fraction of the  stars at lower metallicities and $\log g$, and located in the more distant globular clusters, have their distances overestimated by our method. This seems to be because the SEGUE $\log g$ scale is systematically too low, as attested by simulated samples of SEGUE stars.

Finally, we emphasize that the SEGUE data is very distinct 
from APOGEE in terms of spectral range and resolution. These data also probes much
larger distances than the {\it Hipparcos} based data used in Sec. \ref{hiparcoscomp}.

\subsection{RAVE}
\label{ravecomp}

\begin{figure*}
\centering
\includegraphics[width=0.8\textwidth]{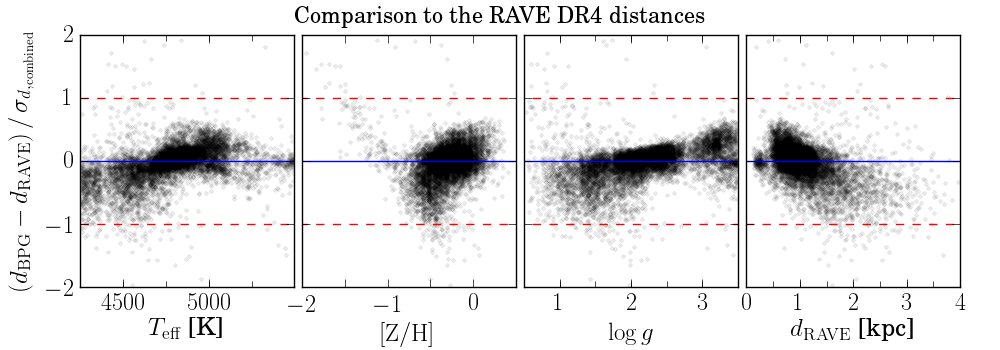}
\caption{\small Comparison of our distance estimates against the RAVE DR4 distances from \cite{Binney2014a} for the high-SNR sample of \citet{Boeche2013}. In each panel the vertical axis shows the absolute distance residuals normalised to the combined uncertainty from both methods, $\sigma_{d_{\mathrm{combined}}} = \sqrt{\sigma(d_{\mathrm{BPG}})^2 + \sigma(d_{\mathrm{RAVE}})^2}$, as a function of the main spectroscopic parameters \Teff, [Z/H] and $\log g$ as well as the RAVE distance, $d_{\mathrm{RAVE}}$.}
\label{ravesample}
\end{figure*}

RAVE collected 
medium-resolution ($R\sim7,500$) CaII triplet spectra of $\gtrsim 400,000$ stars with $9 < I < 13$.
For $> 200,000$ of these stars, \cite{Burnett2011} have been able to derive spectro-photometric distances using spectroscopic parameters from the RAVE pipeline \citep{Kordopatis2011} as well as near infra-red photometry from 2MASS \citep{Skrutskie2006, Cutri2003}. Their method
has provided the background for the work presented in this paper, which is
largely based on the Bayesian approach proposed by \cite{Burnett2011}. More recently,
\cite{Binney2014a} presented an improved version of the method. In this section, we use their distances for a high S/N sample containing $\gtrsim 9000$ giant stars \citep{Boeche2013} to cross-check the
implementation of our method (details in Appendix \ref{raveapp}). The spectroscopic parameters are taken from the RAVE Data Release 4 \citep[DR4;][]{Kordopatis2013}, the photometry is from 2MASS. 

In order to derive distances for RAVE stars, we set $m=K_s$, $\Mabs = M_{K_s}$,  and colors $=\{(J-H), (H-K)\}$. We adopt the $A_{V}$ extinction values that have been computed for individual RAVE objects by \cite{Binney2014a} to deredden the NIR magnitudes of the stars. The conversion between the extinction in the optical and near-IR wavelength range is performed following \cite{Rieke1985}.
We adopted the $\feh$ and $\afe$ estimates from the chemical RAVE pipeline \citep{Boeche2011, Kordopatis2013} to make our metallicity scale more compatible with the stellar models, as in the previous sections. 

Despite the use of a very similar theoretical framework, we note that there are some differences between the way distances are estimated by \cite{Binney2014a} and in this paper. The Padova isochrone grid these authors used \citep{Bertelli2008} is not the same as the one used for this paper, and it was restricted to $\feh \geq -1$. Also, they use a slightly different IMF \citep{Aumer2009}, and apparently they did not use the volume element $s^2ds$ in the density priors presented in Sec. \ref{priors}. They also apply a kinematic correction for their distances \citep{Schoenrich2012a}. From the data side, we have used a slightly different metallicity scale, namely that of the RAVE chemical pipeline instead of the RAVE stellar parameter pipeline, together with a different handling of $\alpha$-enhancement.

Fig. \ref{ravesample} shows the comparison of our distance estimates with the distances from \cite{Binney2014a} for the giant sample analysed in \citet{Boeche2013}. 
For compatibility with our approach described in Sec. \ref{method}, we 
use the $\langle s \rangle$ distance estimate from the calibrated DR4 catalogue from \cite{Binney2014a} in this
comparison, which is the expectation value computed from their model pdf over
distances instead of their corresponding parallaxes. 
Because neither of the distances being compared are supposed to be at the same level of precision as those in our reference samples, we change the way we compare them. Instead of computing the relative residual with respect to any one of the estimates, we normalise the residual by the combined uncertainty from both, which are added in quadrature.
Our distance scale seems to be slightly compressed relative to RAVE DR4: we have systematically larger (smaller) distances than RAVE for stars with $d_{\mathrm{RAVE DR4}} < 1$ ($d_{\mathrm{RAVE DR4}} > 1$) kpc. The effect,
however, is small. The vast majority of the stars have distance residuals well acommodated by the expected uncertainties in the two estimates being compared. 
The mean and rms normalised residuals are $0.00 \pm0.33$,
after a 3-$\sigma$ clipping to eliminate the very few strongly deviant points between the two methods. Looking separately at the two distances regimes, we have 
mean relative residuals of 6\% and -16\%, respectively for $d_{\mathrm{RAVE DR4}} < 1$ and $d_{\mathrm{RAVE DR4}} > 1$ kpc.

\subsection{Effect of different priors}

\begin{figure*}
\begin{center}
\includegraphics[width=\textwidth]{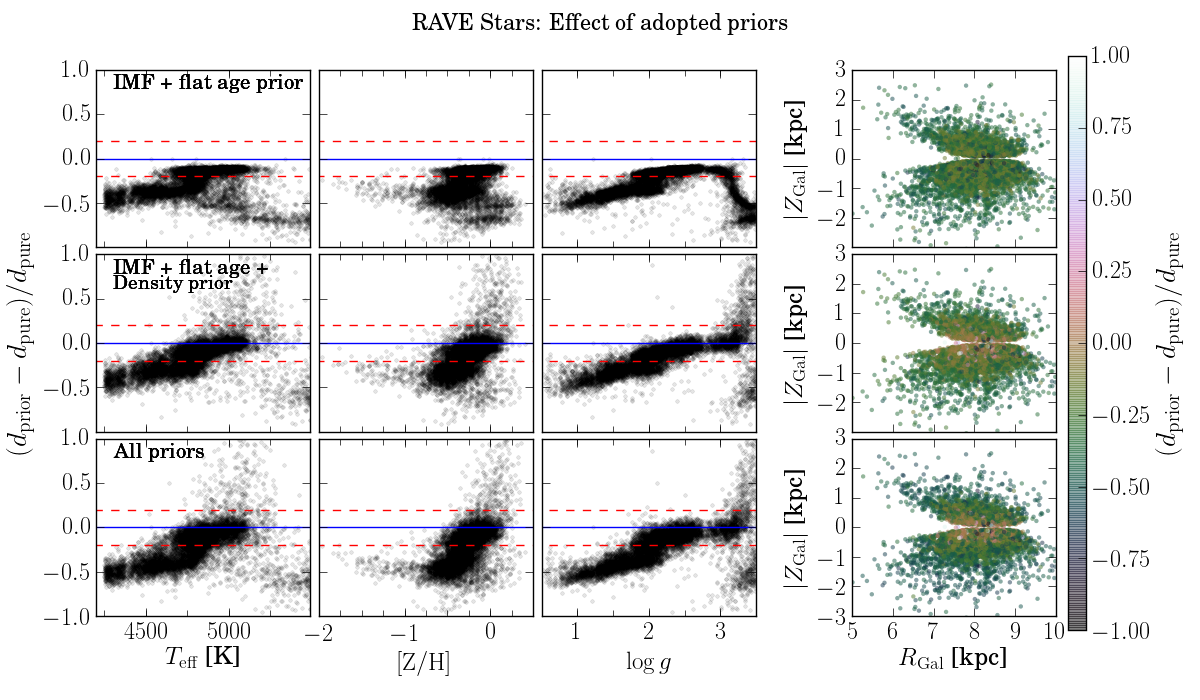}
\caption{\small 
The effect of the adopted priors (Sec. \ref{priors}) on our spectro-photometric distances (BPG), illustrated for the case of the RAVE sample. In each panel, we show the relative distance residuals (priors -- no priors) with respect to the case where no priors were used, as a function of the main stellar parameters \Teff, [Z/H], $\log g$, and as a function of position in the Galaxy (right panels). The {\it top row} shows the results for distances computed using the IMF prior and a flat prior in age instead of the logarithmic one implicit in the spacing of isochrones in $\log \tau$. The {\it middle row} also includes the spatial density priors. Finally, the {\it bottom row} includes the MDF and age priors for the different Galactic components.
}
\label{priortests}
\end{center}
\end{figure*}

In Fig. \ref{priortests}, we assess the effect of the adopted model 
priors on our distance estimates. For this purpose, we use the high quality 
RAVE sample ($\simeq 9000$ stars) presented in Sec. \ref{ravecomp}.

The aim of this exercise is merely to be able to quantify the effect of the adopted priors on a particular stellar population, not the justification of these priors (which is discussed in Sec. \ref{priors}). Note that
for the internal consistency checks discussed in Secs. \ref{simtests}, \ref{foursets} and \ref{sys}, we did not use any priors, which is the appropriate decision for our simulation, but not for real Milky Way stellar populations.

The top row panels in Fig. \ref{priortests} compare distances based on no assumed priors, as used in the simulations, to those using only the IMF prior, as presented in Sec. \ref{priors}, and a correction for a uniform age distribution. The distances in the middle panels include the spatial priors, but still exclude the specific age distributions and MDFs of individual Galactic components presented in Sec. \ref{priors}. The comparison with the distances that also take these later into account is left to the bottom panels in the figure.

It is clear from the figure that inclusion of age and metallicity priors
both tend to systematically reduce the distances of a significant
fraction of the stars. Restricting the thick disc and halo samples to old ages, in particular, prevents models of young luminous stars to be included in
their distance estimates. Similarly, associating disc stars to relatively metal-rich models also tends to reduce their distances. 
A large number of stars are unaffected
by the specific choice of model priors. Using all stars with
estimated distances $s < 6$ kpc (when all priors are included, comprising 
the vast majority of the RAVE high quality sample) we obtain a mean
residual of $5 \%$ and a scatter of $6 \%$ between the {\it all priors}
and {\it spatial priors only} distances. This is, in fact, smaller than
the systematic residuals and scatter found with respect to more 
precisely determined distances,
as discussed in Sec. \ref{validation}.

\section {Summary and conclusions}
\label{conclusion}

We have implemented a code to estimate distances to field stars based on
their measured spectroscopic and photometric parameters. The theoretical
background is very similar to that of \cite{Burnett2010} and 
subsequent papers from the same authors. Briefly, given a set of 
measured quantities, these are compared to model values and, using a
Bayesian approach, the posterior probability distribution function is derived 
for each star. The distance is then determined as the median value of the 
posterior pdf marginalised over the other parameters.

We used simulated stars from PARSEC models as a first validation test of our distances. We showed that distances based on 4 different sets of parameters are well recovered, without significant systematic biases, and statistical uncertainties that scale with the uncertainties of the input parameters. Distances to dwarf stars are more accurately recovered, with $< 1\%$ mean residual and $10\%$ rms errors for a ``high-resolution'' simulated sample; typical errors for giant stars are about twice those values. The method is most sensitive to $\log g$, and starts breaking down if this parameter is removed from the data set, also yielding strong systematic distance errors as a function of bias in this parameter. 

Our simulations also show that the estimated distance uncertainties are in general agreement with the errors, with the uncertainty distribution peaking at $\simeq 10\%$ ($\simeq 20\%$) for dwarf (giant) stars in a mock high-resolution sample, reaching significantly larger values when spectroscopic parameters are less precise. 

We demonstrated that distances of nearby stars with parallaxes
from Hipparcos and precise atmospheric parameters are successfully recovered, with only very small systematic trends, and random errors consistent with the expected combination of errors in the parallaxes and in our distances. The relative random errors are of the order of $20\%$ or less, with $< 1\%$ systematic errors, for nearby FGK dwarf stars. For giants with {\it Hipparcos} parallaxes, the random and systematic residuals in our distance estimates are also as expected based on our mock samples. 

For more distant stars, we have validated the code using giant stars with asterosismic obervations, as well as certain members of well-studied star clusters. In the later case, we compared our distances to the cluster distances taken from the literature. 

Our distances for giant stars in common between APOGEE and CoRoT typically agree within 15\% of each other. A systematic overestimate in our distances in comparison to those inferred from asteroseismology at the level of $\simeq 9\%$ may result from using a more set of comprehensive Galactic priors and a less comprehensive set of photometric measurements than our reference.

We used star clusters included in the SDSS-III SEGUE and APOGEE samples for a final set of validation tests, this time covering a larger range in distances, out to $\simeq 12$ kpc. The distance scatter among individual cluster stars is larger than the distance residuals for the more nearby stars with parallax measurements. For APOGEE clusters, there is a small trend with distance itself, in the sense of underestimating the distances to the more distant cluster stars ($d \gtrsim$ 6 kpc). 

In the case of SEGUE, scatter tends to be higher than APOGEE, consistent with the lower resolution of the spectra. In particular, for a fraction of the low-metallicity giant and subgiant stars that predominate in the more distant globular clusters used, our distances tend to be overestimated, contrary to the trend seen in APOGEE. We have identified the probable origin for this discrepancy as a systematic bias in the SEGUE $\log g$ values. Applying a shift of 0.25 dex to the SEGUE $\log g$ reduces the mean distance residual over individual stars from 14\% to -2\% when all clusters are considered. The scatter is also significantly reduced for the more nearby cluster SEGUE stars. Despite the large scatter observed for a subset of the SEGUE stars, the inferred average distances to the SEGUE clusters are in good agreement with those quoted in the literature.


We have also compared our distances to those for a sample of high-quality RAVE stars, determined with similar approaches to the one presented here \citep{Binney2014a}. Despite the large scatter and a slight systematic trend, our distances for the same sample of red giant stars studied in \cite{Boeche2013} agree, within the expected uncertainties, with those derived by \cite{Binney2014a}.

The validation results are all summarised in Table \ref{summarytable}, where we show the dataset used, the number of stars and the range of reference distances in each, and the mean and rms relative residuals. The table allows a global evaluation of the distances, specially when confronted to those from astrometry, asteroseismology, and well calibrated star clusters. The systematic residuals are usually less than 9\% while the scatter scales with the quality of the input spectro-photometric parameters, with the most distant SEGUE clusters being our worst case.

The code presented and validated in this paper 
is being used by the Brazilian Participation Group on SDSS-III
to derive distances for tens of thousands of stars belonging to the
SEGUE and APOGEE surveys. These surveys are complementary in many ways,
including the directions and Galactic components they probe more efficiently.
When analysed together, provided that reliable distances are available, they
allow a 3D mapping of Galactic structure using different stellar
tracers, determination of chemo-kinematic relations in the Galaxy, 
and ultimately, confrontation of these constraints with models of Galactic
formation and evolution \citep{Minchev2013, Minchev2014b}. Initial science analyses based on these distances are provided by \citet{Junqueira2015} and \cite{Anders2014}.

\begin{acknowledgements}
The authors thank C. Boeche and G. Kordopatis for useful discussions about RAVE.

This research has made use of the WEBDA database, operated at the Institute for Astronomy of the University of Vienna.

Funding for SDSS-III has been provided by the Alfred P. Sloan Foundation, the 
Participating Institutions, the National Science Foundation, and the U.S. 
Department of Energy. The SDSS-III web site is http://www.sdss3.org/.

SDSS-III is managed by the Astrophysical Research Consortium for the 
Participating Institutions of the SDSS-III Collaboration including the 
University of Arizona, the Brazilian Participation Group, Brookhaven National 
Laboratory, University of Cambridge, University of Florida, the French 
Participation Group, the German Participation Group, the Instituto de 
Astrofisica de Canarias, the Michigan State/Notre Dame/JINA Participation 
Group, Johns Hopkins University, Lawrence Berkeley National Laboratory, Max 
Planck Institute for Astrophysics, New Mexico State University, New York 
University, the Ohio State University, University of Portsmouth, Princeton 
University, University of Tokyo, the University of Utah, Vanderbilt University,
 University of Virginia, University of Washington, and Yale University.

Funding for the Brazilian Participation Group has been provided by the 
Minist\'erio de Ci\^encia e Tecnologia (MCT), Funda\c{c}\~ao Carlos Chagas 
Filho de Amparo \`a Pesquisa do Estado do Rio de Janeiro (FAPERJ), Conselho 
Nacional de Desenvolvimento Cient\'{\i}fico e Tecnol\'ogico (CNPq), and 
Financiadora de Estudos e Projetos (FINEP). 

\end{acknowledgements}

\bibliography{FA_library}

\appendix
\section{APOGEE distances and calibration of stellar parameters}\label{apoapp}
For the computation of distances to APOGEE stars, we make use of the ASPCAP stellar parameters bundled in the twelth data release of the Sloan Digital Sky Survey (SDSS DR12; \citealt{Alam2015}). The APOGEE data products contained in this release are described in detail in \citet{Holtzman2015}. While we generally use the stellar parameter calibrations provided by these authors, we have adopted slightly different calibrations for metallicity and surface gravity, which are explained below.

Our APOGEE distances will be released within an SDSS DR12 value-added APOGEE distance catalogue.

{\it Surface gravity for red-clump stars:} \citet{Holtzman2015} use the very accurate and precise surface gravities from the APOGEE-{\it Kepler} sample \citep{Pinsonneault2014} to calibrate out systematic shifts in the surface gravities delivered by ASPCAP. They demonstrate (their Fig. 4) that these systematics are different for red-giant branch (RGB) and core-He burning red-clump stars (RC). The offset between both groups is about 0.2 dex. However, the authors only provide a surface gravity calibration for RGB (their formula 3), and for RC stars defer the reader to the APOGEE RC catalogue described by \citet{Bovy2014} who provide very precise distances to these stars. Because we wanted to test our distance code on {\it all} APOGEE giant stars, we have separately fit the calibration relation for the stars contained in the RC catalogue:
$$\log g_{\mathrm{RC}} = \log g_{\mathrm{calib}} - (7.21\cdot 10^{-5}\cdot(T_{\mathrm{eff}}[\mathrm{K}]-4400)+0.129) $$

{\it Metallicities in the super-solar regime:} ASPCAP metallicities have been calibrated using open and globular cluster metallicities from the literature. Fig. 6 of \citet{Holtzman2015} shows the result of the ''external`` metallicity calibration: in the metal-poor regime, the raw ASPCAP metallicities are up to 0.3 dex too high compared to the literature, while the values are similar at solar metallicities. The authors have therefore opted to fit a second-order polynomial to the data (their formula 6). But because there are very few calibration clusters above solar metallicity, the quadratic calibration relation may be overfitting the data at the metal-rich end, and therefore overestimate metallicities in this regime. We have thus opted to use the calibrated values {\it only when} [Z/H]$<0.0$, and to otherwise use the uncalibrated values.

\section{SEGUE distances}\label{segueapp}

For the computation of SEGUE distances, we use the results of the SEGUE Stellar Parameter Pipeline (SSPP; \citealt{Lee2008, Lee2008a, Lee2011, Smolinski2011}) for a clean sample of G dwarfs (details will be provided in Brauer 2015).
From SDSS DR8 to DR9, it underwent a couple of modifications providing improved stellar parameters for all SEGUE objects within the scope of DR9. In particular,  there are some obvious systematic shifts in the individual parameters when
comparing DR8 and DR9 SSPP values for the same objects.

The most significant deviation is observed for the surface gravity
showing a clear overall decrease of the parameter by 0.2 dex to 0.3 dex to lower values. This finding is in agreement with the fact that the DR9 surface gravity is in general lower by about $\sim 0.2$ dex (Rockosi et al. 2015, in prep.). Possible explanations for the decrease are the following: (1) the surface gravity estimates from the re-analysis of the high-resolution spectra that are used to calibrate the DR9 SSPP results are in general lower by 0.13 dex than those from the older high-resolution analysis, (2) for some of the $\log g$ estimation methods utilised by the SSPP, new synthetic spectral grids were used, and (3) the line index methods \texttt{MgH} and \texttt{CaI2} were not considered any
longer -- which may have had a substantial effect since those estimators produced higher $\log g$ values compared to others. The overall lower gravity is certainly caused by a combination of the above facts.

Besides the shift in surface gravity, the temperatures in DR9 tend to be systematically larger than in DR8 by $\sim 50$ K. This temperature shift occurs due to the IRFM temperature scale being utilised to calibrate the SSPP DR9 temperature estimators.

While the metallicity scale did not change significantly, the [$\alpha$/Fe] abundances shifted to higher values by about 0.1 dex.

{\it $\log g$ calibration:} As shown in Sec. \ref{systematics}, accurate absolute surface gravity values are crucial for computing spectro-photometric distances. Any systematic under- or overestimation of the surface gravity will lead to significant under- or overestimation of our distances.
Hence, as justified in Sec. \ref{seguecomp}, we decided to calibrate
the SEGUE DR9 surface gravity values by applying a shift of 0.25 dex and thus increasing the $log g$ of each object in the G-dwarf sample by the same amount.
Apart from this modification, we use the SEGUE DR9 values delivered by the SSPP.

\section{RAVE distances}\label{raveapp}

In this work, we have used the RAVE giant sample assembled by \citet{Boeche2013}. It comprises stars with the highest quality spectra and abundances. The following quality criteria have been applied: (1) only spectra with a signal-to-noise ratio of at least 60 were selected on which the RAVE pipeline converged to a
single point of the parameter space (Flag $\tt{Algo\_Conv} = 0$), (2) the chemical pipeline converged ($\chi^2 < 1000$), (3) the number of defective pixels along the spectrum is small (\texttt{frac} $> 0.99$), and (4) every object that is not classified as a normal star according to the morphological classification
described in \citet{Matijevivc2012} was excluded. 
Like \citet{Binney2014a}, we use the stellar parameters from
DR4 \citep{Kordopatis2013}. The applied cuts to select RAVE giants are 0.5 dex $< \log g <$ 3.5 dex and 4000 K $< \Teff <$ 5500 K.
This ensures that only cool giants are selected and problems with the grid limits of the automated pipeline are avoided by setting the lower limit in $\log g$. The resulting RAVE giant sample comprises 9098 stars.

We will also provide distances for the full RAVE DR4 sample in the near future.
\end{document}